\documentclass[%
 reprint,
 amsmath,amssymb,
aps,
pre,
nofootinbib]{revtex4-2}

\usepackage[T1]{fontenc}
\usepackage{graphicx}
\usepackage{dcolumn}
\usepackage{bm}

\usepackage{amsmath,scalerel}	
\usepackage{amssymb}	
\usepackage{txfonts}
\usepackage[bb=libus]{mathalpha}
\usepackage{bbold}
\usepackage{relsize}
\usepackage{mathtools}
\usepackage{bigints}
\usepackage{sidecap}
\usepackage{caption,subcaption}
\usepackage{hyperref}

\usepackage[nodisplayskipstretch]{setspace}
\usepackage[usenames,dvipsnames]{color}
\usepackage{floatrow}
\usepackage{soul}
\usepackage{comment}
\usepackage[normalem]{ulem}

\usepackage{macros_ub}
\graphicspath{{./}{figures/}}

\begin{document}

\preprint{APS/123-QED}

\title{Universal power-law distribution functions in an electromagnetic kinetic plasma: implications for the inverted temperature profile in the solar corona}



\author{Uddipan Banik}
\email{uddipanbanik@ias.edu}
\affiliation{Department of Astrophysical Sciences, Princeton University, 112 Nassau Street, Princeton, NJ 08540, USA\\Institute for Advanced Study, Einstein Drive, Princeton, NJ 08540, USA\\Perimeter Institute for Theoretical Physics, 31 Caroline Street N., Waterloo, Ontario, N2L 2Y5, Canada}
\author{Amitava Bhattacharjee}%
\email{amitava@princeton.edu}
\affiliation{Department of Astrophysical Sciences, Princeton University, 112 Nassau Street, Princeton, NJ 08540, USA}

%




\date{\today}

\begin{abstract}

We develop a self-consistent quasilinear theory for the relaxation of electromagnetic kinetic plasmas, and demonstrate that the mean distribution functions of both electrons and ions tend to relax to a universal $v^{-5}$ tail. Large-scale electromagnetic (EM) fields efficiently accelerate the unscreened, fast particles but not the screened, slow ones. This non-thermal tail may arise in the solar corona from EM turbulence despite collisions, allowing suprathermal particles to escape the sun's gravity (velocity filtration) and inverting the temperature $(T)$ profile with $T$ rising to $10^6$ K.

\end{abstract}

\maketitle


\paragraph*{Introduction\textemdash}

It is textbook wisdom that a collisional plasma relaxes to a Maxwellian distribution function (DF) in the steady state, as guaranteed by Boltzmann’s H-theorem. In contrast, a collisionless plasma, for which no H-theorem exists, can accommodate a denumerably infinite number of possible steady state distributions that are functions of the constants of motion. Nevertheless, kinetic (collisionless or weakly collisional) plasmas observed in nature exhibit a tendency to relax to power-law DFs with a Maxwellian core at lower energies and a power-law tail at higher energies. To be specific, a particular power-law DF, $f_0(v)\sim v^{-5}$, where $v$ is the particle velocity, with a corresponding $E^{-2}$ energy ($E$) distribution \citep[][]{Fisk.Gloeckler.14}, has been observed in the solar wind, in the inner heliosphere by the Ulysses and ACE spacecrafts \citep[][]{Fisk.Gloeckler.12} and in the heliosheath by Voyager \citep[][]{Krimigis.etal.77,Stone.etal.77}. A theory has been proposed for the universality of such DFs \citep[][]{Fisk.Gloeckler.14}, but the claim has been questioned on both theoretical and observational grounds \citep[][]{Jokipii.Lee.10}. Some of the criticism stems from the fact that the observed DFs show a degree of variability, either due to collisions or a variety of plasma processes not considered in the theory.

Recently, we have developed a fully \textit{self-consistent} treatment of electrostatic plasmas, which shows that there is indeed an attractor DF, $f_0(v)\sim v^{-5}$, that a collisionless plasma relaxes to under large-scale electric field fluctuations. In contrast, both \citep[][]{Fisk.Gloeckler.14} and \citep[][]{Jokipii.Lee.10} were test-particle treatments that did not include the self-consistent Maxwell's equations. Our theory suggests that the plasma asymptotically evolves into this power-law steady state, but whether it is measured by \textit{in situ} spacecrafts might depend on the collisional age of the plasma and electromagnetic (EM) effects. The development of this power-law tail in a collisionless plasma subject to electrostatic turbulence has been recently demonstrated by a particle-in-cell (PIC) simulation \citep[][]{Ewart.etal.25}.   

Non-Maxwellian DFs are claimed to play a significant role in the unsolved problem of solar coronal heating, i.e., the origin of temperature inversion as one moves outwards from the Sun's surface to the corona. In such cases, heat appears to move from regions of low temperatures to high, in apparent violation of the laws of thermodynamics. An underlying physical mechanism, described as velocity filtration \citep[][]{Scudder.92} (gravity acts as a filter and lets the suprathermal particles escape outwards), depends critically on the assumption that there exists a non-Maxwellian electron (or ion) DF at the base of the corona. While this assumption leads to predictions that appear to be consistent with a significant body of observations in the heliosphere (see the monograph by Meyer-Vernet \citep[][]{Meyer-Vernet.07} for a lucid discussion), the assumption of a non-Maxwellian DF at the coronal base has been questioned because collisional effects at the base are usually significant and are expected to produce Maxwellian-like DFs that do not exhibit the velocity filtration effect. Specifically, theory and idealized kinetic simulations \citep[][]{Anderson.94,Landi.Pantellini.01} demonstrate that, in the presence of collisions, unless one assumes rather shallow power-law tails $(\kappa < 4)$ at the coronal base, it is hard to account for the heat flux and observed temperature inversion. 

In this Letter, we develop a general quasilinear theory (QLT) \citep[c.f.][for a modern review of quasilinear theory]{Diamond_Itoh_Itoh_2010} for the relaxation of electromagnetically driven plasmas, accounting for the simultaneous, self-consistent evolution of multiple species (electrons and ions) in a bath of kinetic plasma turbulence, using the Balescu-Lenard (BL) framework and including the effects of weak binary collisions. We demonstrate the crucial role of the self-consistent EM fields in producing a universal $v^{-5}$ tail in the DF and $E^{-2}$ tail in the energy distribution. This universal scaling persists despite the presence of weak collisions.  The Debye shielding of large-scale fields (on super-Debye scales) occurs through the dielectric tensor governing plasma waves, resulting in the suppression of particle heating at low velocities. The universal scaling manifests itself in both electrons and ions, but as discussed later in the paper, their DFs exhibit interesting differences. Our theoretical treatment demonstrates that in the presence of EM turbulence, the requirements for the power-law at the coronal base are well within range of what is generally observed, fixing some of the fundamental challenges confronting the elegant velocity filtration model as a contender for resolving the coronal heating problem.    

\paragraph*{Relaxation theory for weakly collisional plasmas\textemdash} A plasma is characterized by the DF or phase space ($\bx,\bv$) density of particles of the ${\rms}^{\rm th}$ charged species, $f_\rms(\bx,\bv,t)$. The governing equations for a weakly collisional plasma are the Boltzmann-Maxwell equations. The Boltzmann equation,

\begin{align}
&\frac{\partial f_\rms}{\partial t} + \bv\cdot{\bf \nabla} f_\rms + \frac{q_\rms}{m_\rms}\,{\bf \nabla}_\bv f_\rms \cdot \left[\left(\bE^{(\rmP)}+\bE\right) + \frac{\bv}{c}\times \left(\bB^{(\rmP)}+\bB\right)\right] \nonumber\\
&= C\left[f_\rms\right],
\label{Vlasov_eq}
\end{align}
evolves the DF of each charged species (electrons and ions). Here, $q_\rms$ and $m_\rms$ are the electric charge and mass of each species, and $\bE$ and $\bB$ are the self-generated electric and magnetic fields, sourced by the DF via Maxwell's equations,

\begin{align}
&\nabla\cdot\bE = {4 \pi} \sum_{\rms} q_\rms \int \rmd^3 v\, f_\rms,\nonumber\\
&\nabla \times \bB = \frac{4\pi}{c} \sum_{\rms} q_\rms \int \rmd^3 v\, \bv f_\rms + \frac{1}{c}\frac{\partial \bE}{\partial t},\nonumber\\
&\nabla\cdot\bB = 0,\quad \nabla\times\bE = -\frac{1}{c}\frac{\partial \bB}{\partial t}.
\label{Maxwell_eqs}
\end{align}
We assume charge neutrality in equilibrium, i.e., the total equilibrium charge density is zero. $\bE^{(\rmP)}$ and $\bB^{(\rmP)}$ are perturbing electric and magnetic fields that are sourced by agents (plasma or otherwise) ``external" to our system. Since the plasma is an open system, it is nearly impossible to switch off such fields. Background turbulence spontaneously arising from nonlinear wave activity or coherent structures (electrostatic, e.g., Bernstein-Greene-Kruskal (BGK) modes \citep[][]{Bernstein.etal.57} or electromagnetic, e.g., plasmoids \citep[][]{Loureiro.etal.07,Bhattacharjee.etal.09,Comisso.etal.16}), copiously present within the solar plasma, can also act as part of this drive. $C[f] = \partial/\partial v_i\left(\calD^{(\rms)}_{ij}\left(\bv\right)\partial f_\rms/\partial v_j + \calD^{(\rms)}_i\left(\bv\right) f_\rms\right)$ is the Balescu-Lenard collision operator, with the diffusion ($\calD^{(\rms)}_{ij}$) and drag ($\calD^{(\rms)}_i$) coefficients given by equation~(\ref{BL_coeff_app}). We do not include an equilibrium guide magnetic field in our analysis. 

The nonlinear Boltzmann-Maxwell equations are difficult to solve in their full generality, and demand the use of perturbation theory to make analytical progress. The linear perturbations about a slowly evolving background are computed in Appendix~\ref{App:LT}. Here we formulate a quasilinear (second order perturbation) theory to describe the evolution of the background due to the interference/coupling of the linear modes. 

\paragraph*{Quasilinear theory\textemdash}

The evolution of the mean DF of each species, $f_{\rms 0} = {\left(2\pi\right)}^3 f_{\rms 2\,\bk=0}/V$, averaged over the volume $V$ of the bulk plasma, can be studied by computing the second order response, $f_{\rms 2\,\bk}$, taking the $\bk \to 0$ limit and ensemble averaging the response over the random phases of the linear fluctuations (see Appendix~A.2 of \citep[][]{Banik.Bhattacharjee.24a}). This yields the following quasilinear equation for each species:

\begin{align}
\frac{\partial f_{\rms 0}}{\partial t} &= -\frac{{\left(2\pi\right)}^3 q_\rms}{m_\rms V} \int \rmd^3 k \, \left<\left(\bE_{\bk}^{\ast} + \frac{\bv}{c}\times \bB_{\bk}^{\ast} \right) \cdot \nabla_{\bv} f_{\rms 1\bk}\right>,
\label{quasilin_resp_eq}
\end{align}
where $\bE_{\bk} = \bE_{1\bk} + \bE_{\bk}^{(\rmP)}$ and $\bB_{\bk} = \bB_{1\bk} + \bB_{\bk}^{(\rmP)}$, and the subscript $1$ denotes linear perturbations. We have used the reality condition, $\bE_{1,-\bk} = \bE_{1\bk}^{\ast}$ (similarly for the other quantities).

Now, we need to make assumptions about the temporal correlation of the perturbing EM fields, $E_{\bk i}^{(\rmP)}(t)$ and $B_{\bk i}^{(\rmP)}(t)$, where the subscript $i$ denotes the $i^{\rm th}$ component. Faraday's law (fourth of Maxwell's equations~[\ref{Maxwell_eqs}]) dictates that the electric and magnetic field perturbations are related by 
$\Tilde{\bB}_{\bk}^{(\rmP)}\left(\omega\right) = \frac{c}{\omega} \left(\bk\times\Tilde{\bE}_{\bk}^{(\rmP)}\left(\omega\right)\right)$. We assume that the perturbing electric field $E_{\bk i}^{(\rmP)}(t)$ is a generic red noise:

\begin{align}
\left<E_{\bk i}^{(\rmP)\ast}(t) E_{\bk j}^{(\rmP)}(t')\right> &= \calE_{ij}\left(\bk\right)\,\calC_t\left(t-t'\right),
\label{white_noise_t}
\end{align}
where $\calC_t$ is the temporal correlation function. For white noise, this is simply $\delta\left(t-t'\right)$, whose Fourier transform is $\calC_{\omega}\left(\omega\right) = 1$.

Substituting the expressions for the linear quantities, $\bE_{\bk}(t)$ and $f_{\rms 1\bk}(\bv,t)$, obtained by performing the inverse Laplace transform of equations~(\ref{lin_resp_eq_app}) in the quasilinear equation~(\ref{quasilin_resp_eq}) above, and using the noise spectrum for the perturbing electric field given in equation~(\ref{white_noise_t}), we obtain a simplified form for the quasilinear transport equation for each charged species,

\begin{align}
\frac{\partial f_{\rms 0}}{\partial t} &= \frac{\partial}{\partial v_i}\left[\left(D_{ij}^{(\rms)}(\bv) +  \calD^{(\rms)}_{ij}(\bv)\right)\frac{\partial f_{\rms 0}}{\partial v_j} + \calD^{(\rms)}_i\left(\bv\right)\, f_{\rms 0} \right],
\label{quasilin_resp_FP_eq}
\end{align}
which is a Fokker-Planck equation with the diffusion tensor $D_{ij}^{(\rms)} + \calD_{ij}^{(\rms)}$ and the drag/friction tensor $\calD_i^{(\rms)}$. $D_{ij}^{(\rms)}$ is sourced by the EM drive, while $\calD_{ij}^{(\rms)}$ and $\calD_i^{(\rms)}$ are the BL coefficients sourced by the internal fields and collisions. 

At long time, after the Landau modes have damped away (assuming that the plasma is stable), the drive diffusion tensor $D_{ij}^{(\rms)}$ is given by (see Appendix~\ref{App:LT} for a discussion of when and why the Landau term can be neglected)

\begin{align}
&D_{ij}^{(\rms)}(\bv) \approx \frac{8\pi^4 q^2_\rms}{m^2_\rms V} \int \rmd^3 k\, {\left[ \varepsilon^{-1}_{\bk}\left(\bk\cdot\bv\right)\, \varmathbb{P}_{\bk}\left(\bk\cdot\bv\right)\, \varepsilon^{-1\dagger}_{\bk}\left(\bk\cdot\bv\right) \right]}_{ij},\nonumber\\
&\varmathbb{P}_{\bk\, ij}\left(\bk\cdot\bv\right) = \frac{k_i v_l\, \calE_{lj}\left(\bk\right) \calC_{\omega}\left(\bk\cdot\bv\right)}{\bk\cdot\bv},
\label{diffusion_tensor}
\end{align}
with the dielectric constant $\varepsilon_{\bk}\left(\bk\cdot\bv\right)$ given by equation~(\ref{eps_k_app}). Its longitudinal component scales universally as $1 - \omega^2_{\rmP e}/{\left(\bk\cdot\bv\right)}^2$ ($\omega_{\rmP e} = \sqrt{4\pi n_e e^2/m_e}$ is the plasma frequency) over a large $v$ range for super-Debye fields $\left(k\lambda_\rmD \ll 1\right)$, independent of the detailed functional form of $f_{\rms 0}$, implying that slower particles are more Debye-shielded. 


\paragraph*{A general transport equation for particle acceleration\textemdash}

To reduce the dimensionality of the problem, we assume

\begin{itemize}
    \item Isotropic EM drive: $\calE_{ij}(\bk) = \calE(k)\,\hat{k}_i\hat{k}_j$
    \item Isotropic DF: $f_{\rms 0}(\bv) = f_{\rms 0}(v)$,
\end{itemize}
with $\hat{k}_i = k_i/k$. The assumption of isotropy is justified in the absence of a strong guide field. The transport equation thus becomes the following one-dimensional equation in $v$:

\begin{align}
&\frac{\partial f_{\rms 0}}{\partial t} = \frac{1}{v^2}\frac{\partial}{\partial v}\left[v^2 \left( \left(D^{(\rms)}(v) + \calD^{(\rms)}_2(v)\right)\frac{\partial f_{\rms 0}}{\partial v} + \calD^{(\rms)}_1(v) f_{\rms 0} \right)\right].
\label{QL_eq}
\end{align}
The drive diffusion coefficient $D^{(\rms)}$ is given by

\begin{align}
D^{(\rms)}(v) &= \frac{32\pi^5 q^2_\rms}{m^2_\rms} \int_0^{\infty} \rmd k\, k^2 \calE(k) \nonumber\\
&\times \int_0^1 \rmd\cos{\theta}\,\cos^4\theta\, \frac{ \calC_{\omega}\left(k v \cos{\theta}\right)}{{\left|\varepsilon_{k\parallel}\left(kv\cos\theta\right)\right|}^2},
\end{align}
where $\theta$ is the angle between $\bk$ and $\bv$, and $\varepsilon_{k\parallel}$, the parallel component of the dielectric tensor, denotes the Debye screening of large-scale EM perturbations. Note that the transverse component does not show up in the dressing of the fields if the fluctuations are isotropic, in which case the particle acceleration is governed by longitudinal electric fields and the EM problem of quasilinear relaxation reduces to an electrostatic one \citep[][]{Banik.Bhattacharjee.24a}. And since magnetic perturbations are always transverse $\left(\bk\cdot \bB_{\rm \bk} = 0\right)$, the acceleration in the isotropic setup is caused by electric field perturbations that are perpendicular to magnetic ones, akin to Fermi \citep[][]{Fermi.49} acceleration.

The diffusion ($\calD^{(\rms)}_2$) and drag ($\calD^{(\rms)}_1$) BL coefficients are given by

\begin{align}
\calD^{(\rms)}_i(v) &= \frac{\omega_{\rmP\rms}^4}{n_\rms} \int \frac{\rmd k}{k} \int_0^1 \rmd\cos{\theta}\,\frac{\cos^2\theta\; \calF_{\rms i}\left(v\cos{\theta}\right)}{{\left|\varepsilon_{k\parallel}\left(kv\cos\theta\right)\right|}^2},
\end{align}
where $i=1,2$, $\calF_{\rms 1}\left(v\cos{\theta}\right) = v f_{\rms 0}\left(v\cos{\theta}\right)$, and $\calF_{\rms 2}\left(v\cos{\theta}\right) = F_{\rms 0}\left(v\cos{\theta}\right)$, $F_{\rms 0}(v) = 2\pi \int \rmd v'\,v' f_{\rms 0}\left(\sqrt{v^2 + v'^2}\right)$ being the one-dimensional DF. In deriving these, we have neglected the cross-terms denoting inter-species interactions. This is justified if one of the species is significantly heavier, as is the case for a hydrogen or helium plasma, since electron-ion interactions scatter electrons off the inert ions, only changing the angular but not the $v$ dependence of their DFs.

\begin{figure}
\centering
\includegraphics[width=0.85\textwidth]{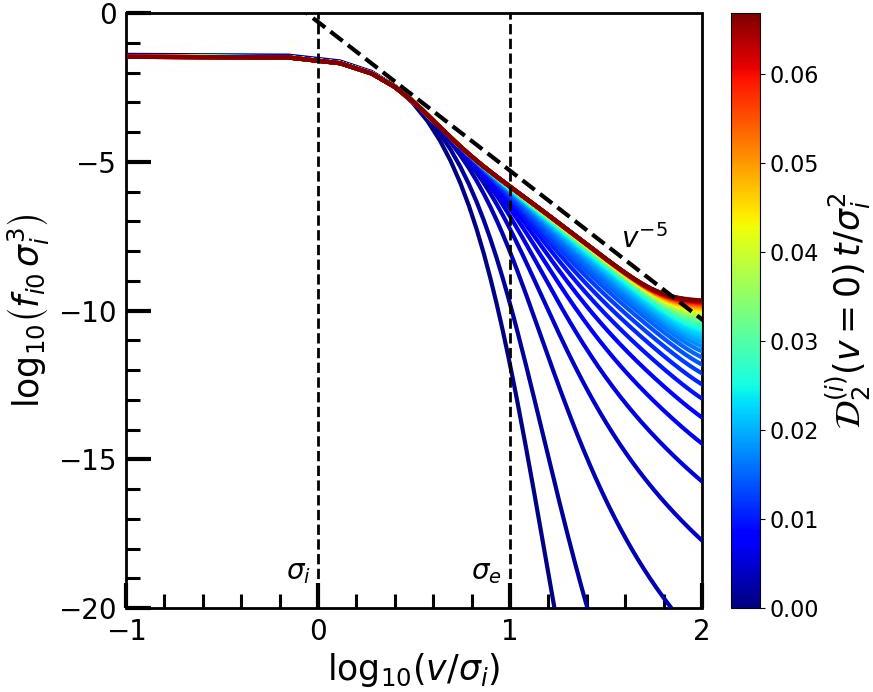}
\caption{Development of the $v^{-5}$ tail in the ion distribution due to large-scale EM turbulence, obtained by solving equation~(\ref{QL_eq}).}
\label{fig:fi_vs_t}
\end{figure}

\begin{figure}
\centering
\includegraphics[width=0.85\textwidth]{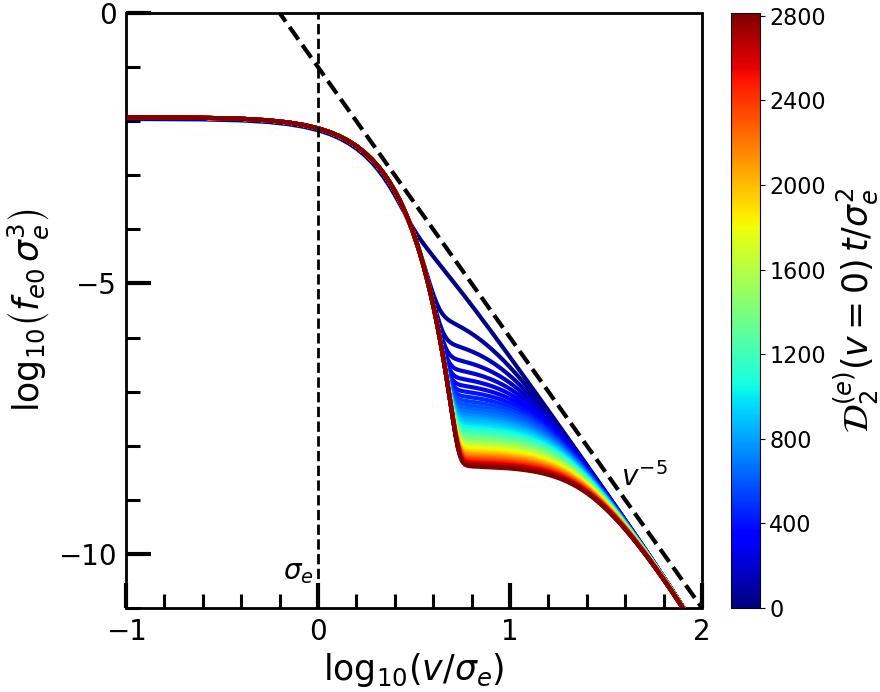}
\caption{Maxwellianization of the electron DF after the EM drive has been switched off and the plasma relaxes due to turbulent and collisional relaxation through the BL equation (equation~[\ref{QL_eq}] without the drive). Note the continued presence of runaway electrons.}
\label{fig:fe_vs_t_after_shock}
\end{figure}

Let us now compare the different coefficients. For simplicity, we assume that the plasma is composed of electrons and ions of a single species. We assume the drive to be a temporal white noise ($\omega_{\rmP e} t_\rmc < 1$ with $t_\rmc$ the noise correlation time) that dominates on super-Debye scales ($k_\rmc \lambda_\rmD \ll 1$ with $k_\rmc$ the characteristic wavenumber). Vlasov turbulence is generally of the form, $\calE(k)\sim k^{-8}$, in 3D \citep[][]{Ewart.etal.25,Nastac.etal.23,Nastac.etal.25,Ginat.etal.25}, which has this property. The internal diffusion (drag) coefficient $\calD_2^{(\rms)}$ ($\calD_1^{(\rms)}$) is constant (scales as $\sim v$) at $v\lesssim \sigma_\rms$ and scales as $\sim v^{-3}$ ($\sim v^{-2}$) at $v\gtrsim \sigma_\rms$ (see Fig.~\ref{fig:D_vs_v} of Appendix~\ref{App:QLT}). For $T_e \sim T_i$, the drive diffusion coefficient  $D^{(\rms)}$ is constant for $v \lesssim \sigma_e = \sqrt{k_\rmB T_e/m_e}$ and scales as $\sim v^4$ for ${\rm max}\left(\sigma_e,\nu_{\rmc \rms}/k_\rmc\right) \lesssim v \lesssim \omega_{\rmP e}/k_\rmc = \sigma_e/k_\rmc\lambda_\rmD$, for super-Debye forcing ($\nu_{\rmc \rms} \approx \omega_{\rmP\rms}\ln\Lambda/\Lambda$ is the collision frequency, $\Lambda = n_e \lambda^3_\rmD$ being the plasma parameter). This is because the dielectric constant $\varepsilon_{k\parallel}$ scales as $\sim {\left(\omega_{\rmP e}/kv\right)}^2$ in this velocity range \citep[][]{Banik.Bhattacharjee.24a}. Physically, this implies that slower particles are more Debye shielded and less accelerated, while the very fast ones are unscreened and readily heated, even in the presence of collisions. Redness of the drive $(\omega_{\rmP e}t_\rmc > 1)$ preserves the $v^4$ behavior of $D^{(\rms)}$ up to $v \sim 1/k_\rmc t_\rmc$ but makes it shallower beyond. The drive diffusion coefficient always exceeds the BL coefficients at large enough $v$, irrespective of the strength of the drive, as long as it predominantly acts on super-Debye scales.

We evolve the electron and ion DFs by numerically integrating equation~(\ref{QL_eq}) for each species, using a flux conserving scheme detailed in Appendix~C.1 of \citep[][]{Banik.Bhattacharjee.24a}, and assuming an isotropic, white noise ($\calC_\omega = 1$) EM drive with $k_\rmc \lambda_\rmD = 10^{-2}$, for which $D^{(s)}(v)\sim v^4$ over a large $v$ range. We plot the resulting ion DF $f_{i 0}$ in Fig.~\ref{fig:fi_vs_t} (the electron DF has similar, albeit faster, evolution since $D^{(e)} \sim D^{(i)} {\left(m_i/Z m_e\right)}^2 \gg D^{(i)}$). 
Evidently, an initial Maxwellian-like $f_{i 0}$ develops a $v^{-5}$ tail. The origin of this tail can be understood by working out the steady state solution to the quasilinear equation~(\ref{QL_eq}):

\begin{align}
&f_{\rms 0}(v) = N\,\exp{\left[-\int\rmd v'{\calD_1^{(\rms)}\left(v'\right)}\Big/\left({\calD_2^{(\rms)}\left(v'\right) + D^{(\rms)}\left(v'\right)}\right)\right]}\nonumber\\
&\times \left[1 + c \int \rmd v' \dfrac{\exp{\left[\int\rmd v''{\calD_1^{(\rms)}\left(v''\right)}\Big/\left({\calD_2^{(\rms)}\left(v''\right) + D^{(\rms)}\left(v''\right)}\right)\right]}}{v'^2\left(\calD_2^{(\rms)}\left(v'\right) + D^{(\rms)}\left(v'\right)\right)}\,\right],
\end{align}
with $N$ a normalization constant and $c$ an integration constant related to the flux. The first term, a zero flux solution, can be shown to be a Maxwellian. The second term, a constant flux solution, is close to a $\kappa$ distribution \citep[][]{Livadiotis.McComas.13,Zhdankin.22a,Zhdankin.22b}, $f_{\rms 0}(v)\sim {(1+v^2/2\kappa\sigma^2)}^{-\left(1+\kappa\right)}$ with $\kappa = 1.5$, which falls off as $\sim \int_v^\infty \rmd v'/{v'}^6 \sim v^{-5}$ at ${\rm max}\left(\sigma_e,\nu_{\rmc \rms}/k_\rmc,v_{\rm crit}\right) \lesssim v \lesssim {\rm min}\left(\omega_{\rmP e}/k_\rmc,1/k_\rmc t_\rmc\right)$ (since $D^{(\rms)} + \calD_2^{(\rms)}\sim v^4$ therein), with $v_{\rm crit}$ the critical velocity beyond which the drive exceeds the internal diffusion (see Appendix~\ref{App:criteria}) and $t_\rmc$ the correlation time of fluctuations of the drive ($\lesssim \omega_{\rmP e}^{-1}$ for white noise). As shown in \citep[][]{Banik.Bhattacharjee.24a}, redness of the drive renders the power-law fall-off of $f_{\rms 0}$ shallower than $v^{-5}$ for $1/k_\rmc t_\rmc < v < \omega_{\rmP e}/k_\rmc$ (for $\omega_\rmP t_\rmc > 1$). Beyond $v\sim \omega_{\rmP e}/k_\rmc$, $f_{\rms 0}$ is super-exponentially truncated since $D^{(\rms)}$ falls off with $v$ therein \citep[][]{Banik.Bhattacharjee.24a}, which renders all moments finite.

As shown in Fig.~(\ref{fig:fi_vs_t}), the DF features a Maxwellian core with a suppressed $v^{-5}$ tail at intermediate times. The log-slope approaches (is steeper than) $-5$ at large (intermediate) $v$. The tail grows in strength, and $f_{\rms 0}$ approaches a $\kappa$ distribution with $\kappa=1.5$ at long time. The energy ($E = m_s v^2/2$) distribution $N_\rms(E) = g(E)f_{\rms 0}(E)$, with $g(E)\sim v$ the density of states, approaches $v^{-4}\sim E^{-2}$ (see Appendix~\ref{App:entropy} for a discussion of entropy-based arguments \citep[][]{Livadiotis.McComas.13,Zhdankin.22a,Zhdankin.22b,Ewart.etal.22,Ewart.etal.23,Ewart.etal.25} to obtain this). Once the EM drive is switched off, the electron DF Maxwellianizes via turbulent and collisional relaxation (see Fig.~\ref{fig:fe_vs_t_after_shock}), and does so much faster than the ion DF. However, the $v^{-5}$ tail persists, albeit suppressed, at high $v$, indicating the continued presence of runaway particles. We show in Appendix~\ref{App:criteria} that the ratio of the amplitude of the drive to that of the internal turbulence must exceed a critical threshold for the tail to form.

\paragraph*{Application to the solar corona\textemdash}\label{sec:corona}

\begin{figure}
\centering
\includegraphics[width=0.75\textwidth]{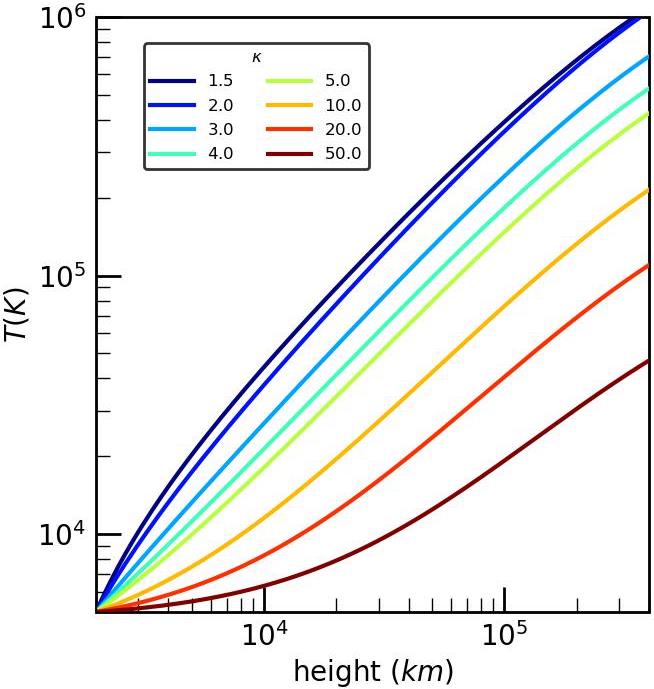}
\caption{Temperature profile in the corona as a function of height above the photosphere, obtained from equation~(\ref{rho_T_profile})}
\label{fig:T_corona}
\end{figure}

So far, our treatment of the plasma has been local, i.e., the coarse-graining scale has been assumed to be smaller than macroscopic scales such as the gravitational scale height $h$. To study the behavior of the plasma on scales $\sim h$, we must take into account the effect of the attractive gravitational potential. To describe the behavior of macroscopic quantities in the weakly collisional plasma of the solar corona, we can construct an effective fluid model from the kinetic description. Our goal here is to focus on the simplest, qualitative deviation from the original velocity filtration model \citep[][]{Scudder.92} that serves as an Occam's razor approach towards resolving the coronal heating problem.


The Vlasov equation dictates that, under the sun' gravity, both the electron and ion DFs in the corona become $f_{\rms 0}(v,r)\sim {\left[1 + {(v^2 + \Phi(r) - \Phi(r_\rmb))}/2\kappa\sigma^2_\rms\right]}^{-\left(1+\kappa\right)}$ \citep[][]{Scudder.92}, with a particular preference for $\kappa = 1.5$, where $r$ is the radial distance from the center, $r_\rmb$ the radius of the coronal base, and $\Phi(r) \approx -G M_{\odot}/r$ the gravitational potential ($M_\odot$ is the solar mass). The preferred value of $\kappa = 1.5$ lies within a plausible range for protons and heavy ions in the fast solar wind (see \citep[][]{Pierrard.Lazar.10} and references therein).  The corresponding density $\rho(r) = 4\pi \sum_{\rms}\int \rmd v\, v^2 f_{\rms 0}(v,r)$ and pressure $p(r) = 4\pi \sum_{\rms}\int \rmd v\, v^4 f_{\rms 0}(v,r)$ are given by equations~(\ref{p_rho_app}). The profiles for the density and temperature $T = T_e + T_i = \left(\mu m_\rmP/k_\rmB\right)\left(p/\rho\right)$ ($\mu$ is the mean atomic weight and $m_\rmP$ is the proton mass) thus turn out to be

\begin{align}
\rho(r) &\approx \rho(r_\rmb)\,{\left[1 + \frac{\mu G M_\odot m_\rmP}{2\kappa k_\rmB T(r_\rmb)}\left(\frac{1}{r_\rmb} - \frac{1}{r}\right)\right]}^{-\left(\kappa-1/2\right)},\nonumber\\
T(r) &\approx T(r_\rmb) + \frac{\mu G M_\odot m_\rmP}{2\kappa k_\rmB} \left(\frac{1}{r_\rmb} - \frac{1}{r}\right).
\label{rho_T_profile}
\end{align}
The same can be obtained by solving the equation for hydrostatic equilibrium,

\begin{align}
\frac{\rmd p}{\rmd r} + \rho\frac{\rmd \Phi}{\rmd r} = 0,
\end{align}
with a macroscopic equation of state (EOS), $p = p\left(\rho\right)$, obtained from equations~(\ref{p_rho_app}), assuming the plasma to be at rest on an average. For $\kappa > 1.5$, the EOS is polytropic, i.e., $p \propto \rho^\gamma$ with $\gamma = \left(\kappa - 3/2\right)/\left(\kappa - 1/2\right)$ (as noted by \citep[][]{Scudder.92}), while, for $\kappa = 1.5$, it is a much milder $p \propto \ln{\left[{\left(v_{\rm max}/\sigma_e\right)}\,\rho/\rho(r_\rmb)\right]}$ with $v_{\rm max} \approx \omega_{\rmP e}/k_\rmc$ (see Appendix~\ref{App:EOS}).

The temperature (density) is therefore an increasing (decreasing) function of $r$ in the corona. This is because the collisionless nature of the plasma yields a softer than isothermal EOS. As $\rho$ decreases outwards due to gravity, $p$ does so too but not quite enough, so that $T\sim p/\rho$ rises outwards. As $\kappa$ approaches $1.5$, the EOS gets softer and the temperature rises more dramatically. Assuming $T(r_\rmb) = 5000$ K, $\mu = 0.5$, $M_\odot = 2\times 10^{30}$ kg, and $r_\rmb$ equal to the solar radius $R_\odot = 7\times 10^5$ km, the corresponding $T$ profile is computed as a function of the height above the photosphere, and plotted for different values of $\kappa$ in Fig.~\ref{fig:T_corona}. While $T$ increases outwards in all cases, only for $\kappa \approx 1.5$ it resembles the observed profile with $T \sim 10^6$ K in the outer corona. For a Maxwellian-like (large $\kappa$) DF, the EOS is nearly isothermal and $T$ does not rise as much. 

Gravity allows high energy suprathermal particles to preferentially escape outwards, which heats up the outer corona to $\sim 10^6$ K. The fact that the virial temperature of the sun, $\mu G M_\odot m_\rmP/k_\rmB R_\odot$, is $\sim 10^7$ K, indicates that gravity has a crucial role to play. Ultimately, it is the $v^{-5}$ non-thermal tail that arises at the coronal base from large-scale EM turbulence (despite the presence of collisions), which enables the velocity filtration to take effect. This alleviates the concerns raised by \citep[][]{Anderson.94,Landi.Pantellini.01} about the interference of collisional effects with velocity filtration. Maxwellianization through collisions cannot suppress the suprathermal tail as long as large-scale EM turbulence keeps stirring the plasma.

\paragraph*{Conclusion\textemdash}\label{sec:conclusion}

We have developed a novel theory for non-thermal particle acceleration in electromagnetically driven, multi-species, kinetic plasmas, which includes the effects of self-consistency, not included in earlier test-particle treatments \citep[][]{Parker.65,Jokipii.Lee.10}. Using the framework of quasilinear theory, we have obtained a general transport equation for the mean DF of each charged species under the action of the drive as well as internal turbulence and weak collisions. We demonstrate the crucial role of the self-consistent EM field in spawning a universal $v^4$ dependence of the quasilinear diffusion coefficient and the resultant $v^{-5}$ non-thermal tail in the DF and $E^{-2}$ tail in the energy distribution. Debye shielding of large-scale (super-Debye) fields occurs through the dielectric tensor, resulting in the suppression of particle heating at low velocities. The very fast particles are unscreened and readily accelerated, unaffected by collisions. While the tail is present in both ions and electrons, it develops much quicker in electrons than in the heavier ions, but is also quenched more rapidly (yet survives as a runaway population) in electrons after the drive is switched off, due to relatively quicker Maxwellianization through turbulent and collisional relaxation. The DF of a partially relaxed or post-perturbation plasma shows steeper fall-offs, which could explain the variability of the power-law tail in the solar wind data. The above power-law DF, potentially generated by large-scale EM turbulence at the base of the weakly collisional solar corona, is sufficient to enable the preferential escape of suprathermal particles to the outer corona (velocity filtration), inverting the temperature profile and raising $T$ up to $\sim 10^6$ K. In future work, we intend to extend our quasilinear analysis to magnetized plasmas (strong guide field) and describe particle acceleration therein, using a generalized (self-consistent) version of the Parker transport equation \citep[][]{Parker.65}.

\begin{acknowledgments}
The authors are thankful to Srijan Das, Mihir Desai, Robert Ewart, Barry Ginat, Nuno Loureiro, Michael Nastac, Alex Schekochihin, Anatoly Spitkovsky, Dmitri Uzdensky and Vladimir Zhdankin for stimulating discussions and valuable suggestions. The first author (UB) is also thankful to the Plasma Science Fusion Center at MIT for providing an intellectually stimulating environment. This research is supported by the National Science Foundation Award 2206607 at the Multi-Messenger Plasma Physics Center (MPPC), and Princeton University.
\end{acknowledgments}

\appendix

\section{Perturbation theory}\label{App:LT}

If the strength of the electrostatic potential, $\Phi^{(\rmP)} = -\int \bE^{(\rmP)}\cdot \rmd \bx$, is smaller than $\sigma^2_\rms$, where $\sigma_\rms$ is the velocity dispersion or thermal velocity of the $\rms^{\rm th}$ species in the unperturbed near-equilibrium system, then the perturbation in $f_\rms$ can be expanded as a power series in the small perturbation parameter, $\epsilon \sim \left|\Phi^{(\rmP)}\right|/\sigma^2_\rms$, i.e., $f_\rms = f_{\rms 0} + \epsilon f_{\rms 1} + \epsilon^2 f_{\rms 2} + ...\,$; $\bE$ and $\bB$ can also be expanded accordingly, assuming $\bE^{(\rmP)}$ and $\bB^{(\rmP)}$ to be $\calO(\epsilon)$. We perform a Fourier transform with respect to $\bx$ and Laplace transform with respect to $t$ of $f_{\rms 1}$, $\bE_1$, $\bE^{(\rmP)}$, $\bB_1$ and $\bB^{(\rmP)}$ in the linearized Vlasov-Maxwell equations, to derive the response of the system order by order. 

\subsection{Linear theory}\label{App:LT}

The Fourier-Laplace coefficients of the linear response for a weakly collisional plasma, in the large mean free path limit of $\nu_{\rmc\rms} \ll k\sigma_\rms$, where $\nu_{\rmc\rms} \sim \omega_{\rmP \rms} \ln\Lambda/\Lambda$ is the collision frequency, can be summarized as follows:

\begin{align}
&\Tilde{f}_{\rms 1\bk}(\bv,\omega) = -\frac{iq_\rms}{m_\rms} \, \frac{\left(\Tilde{\bE}^{(\rmP)}_{\bk}(\omega) + \Tilde{\bE}_{1\bk}(\omega)\right)\cdot {\partial f_{s0}}/{\partial \bv}}{\omega - \bk\cdot\bv} + \frac{i f_{\rms 1 \bk}\left(\bv,0\right)}{\omega - \bk\cdot\bv},\nonumber\\
&\Tilde{E}_{\bk i}(\omega) = \Tilde{E}^{(\rmP)}_{\bk i}(\omega) + \Tilde{E}_{1\bk i}(\omega) = \varepsilon^{-1}_{\bk\, ij}(\omega)\,\left[{\Tilde{\bE}^{(\rmP)}_{\bk j}(\omega)} + g_{\bk j}(\omega)\right],
\label{lin_resp_eq_app}
\end{align}
with the dielectric tensor ${\varepsilon}_{\bk}(\omega) = {\rm diag}\left(\varepsilon_{\bk \perp},\varepsilon_{\bk \perp},\varepsilon_{\bk \parallel}\right)$ given by

\begin{align}
&\varepsilon_{\bk \parallel}(\omega) = 1 + \sum_{\rms}\frac{\omega^2_{\rmP \rms}}{k^2} \int \rmd^3 v\, \frac{\bk\cdot{\partial f_{\rms 0}}/{\partial \bv}}{\omega - \bk\cdot\bv},\nonumber\\
&\varepsilon_{\bk \perp}(\omega) = 1 - 
\frac{\omega}{c^2k^2-\omega^2} \sum_s {\omega^2_{\rmP \rms}} \int \rmd^3 v\, \frac{v_{\perp}{\partial f_{\rms 0}}/{\partial v_\perp}}{\omega - \bk\cdot\bv},
\label{eps_k_app}
\end{align}
and the vector $g_{\bk}(\omega) = \left(g_{\bk\perp},g_{\bk\perp},g_{\bk\parallel}\right)$, corresponding to the initial perturbation, given by

\begin{align}
&g_{\bk\parallel}(\omega) = \frac{4 \pi}{k} \sum_{\rms} {n_\rms q_\rms} \int \rmd^3 v\, \frac{f_{\rms 1\bk}\left(\bv,0\right)}{\omega - \bk\cdot\bv},\nonumber\\
&g_{\bk\perp}(\omega) = -\frac{4 \pi\omega}{c^2 k^2 - \omega^2} \sum_{\rms} {n_\rms q_\rms} \int \rmd^3 v\, \frac{v_\perp f_{\rms 1\bk}\left(\bv,0\right)}{\omega - \bk\cdot\bv}.
\label{g}
\end{align}
Here $\omega_{\rmP \rms} = \sqrt{4\pi n_\rms q^2_\rms/m_\rms}$ is the plasma frequency (or the frequency of ion waves), $n_\rms$ being the number density of the charged species. The subscript $\bk$ stands for the Fourier transform in $\bx$, and the tilde represents the Laplace transform in $t$. Without loss of generality, we have assumed $\bk = k\,\hat{\bz}$ with the perpendicular space spanned by $\hat{\bx}$ and $\hat{\by}$, and $u_{\perp}$ ($v_{\perp}$) denoting the component of $\bu$ ($\bv$) perpendicular to $\bk$, i.e., either $u_x$ ($v_x$) or $u_y$ ($v_y$). We have assumed that the DF $f_{\rms 0}$ is isotropic in $\bu$, i.e., $f_{\rms 0}(\bu) = f_{\rms 0}(u)$, such that $\partial f_{\rms 0}/\partial \bu = \partial f_{\rms 0}/\partial u \, \hat{\bu}$ and therefore the magnetic Lorentz force is zero at linear order.

The dielectric tensor represents (i) the longitudinal polarization of the medium that manifests as Debye shielding/screening of the electric field parallel to $\bk$ and (ii) the transverse EM or light waves, perpendicular to $\bk$, that are modulated by plasma oscillations. The real part of the longitudinal component $\varepsilon_{\bk \parallel}$ universally scales as $1 - \omega^2_{\rmP e}/\omega^2$ for $\omega \gg k\sigma_e$, independent of the detailed functional form of $f_{\rms 0}$. 

The zeros of the dielectric tensor denote the Landau modes or waves that oscillate and (Landau) damp. The electron Langmuir waves ($\omega^2 \approx \omega^2_{\rmP e} + 3 k^2\sigma^2_e$) and the ion Langmuir and ion acoustic waves ($\omega^2 \approx k^2 c^2_\rms/(1+k^2\lambda^2_{\rmD e})$), with $\omega_{\rmP e} = \sqrt{4\pi n_e e^2/m_e}$ the electron plasma frequency ($n_e=$ electron number density), $c_\rms = \sqrt{k_\rmB Z T_e/m_i}$ the sound speed ($T_e$ = electron temperature, $Z$ the atomic number of the dominant ionic species) and $\lambda_{\rmD e}=\sigma_e/\omega_{\rmP e}$ the electron Debye length, emerge from the zeros of $\varepsilon_{\bk \parallel}\left(\omega\right)$ and are longitudinal waves, with electric field oscillations parallel to $\bk$, that get Landau damped via wave-particle interactions. The transverse light waves, with EM oscillations perpendicular to $\bk$, arise from the zeros of $\varepsilon_{\bk \perp}\left(\omega\right)$, follow the dispersion relation $\omega^2 \approx c^2 k^2 + \sum_\rms \omega^2_{\rmP\rms}\left(1+\sigma^2_\rms/c^2\right)$, and do not undergo Landau damping due to the absence of superluminal particles.

In the small mean-free-path limit $(\nu_{\rmc e} \gg k\sigma_\rms)$, the response can still be described by equations~(\ref{lin_resp_eq_app}) for $v \gg \nu_{\rmc e}/k$. The functional form of the dielectric tensor is, however, modified. Assuming a simplified Lenard-Bernstein form of the collision operator \citep[][]{Lenard.Bernstein.58}, where $D_1^{(\rms)}(v) = \nu_{\rmc\rms} v$ and $D_2^{(\rms)}(v) = \nu_{\rmc\rms} \sigma^2_\rms$ (the Balescu-Lenard coefficients scale the same way at $v \lesssim \sigma_\rms$), the longitudinal component, in the limit of $\omega \gg k\sigma_e$, can be written as follows \citep[][]{Lenard.Bernstein.58,Banik.Bhattacharjee.24b}:

\begin{align}
\varepsilon_{\bk\parallel}(\omega) \approx 1 - \frac{\omega^2_{\rmP e}}{\omega^2}\dfrac{1}{1 + \dfrac{i\nu_{\rmc e}}{\omega}}.
\end{align}
Here we have neglected the sub-dominant ion term. For $\omega \gg \nu_{\rmc e} \gg k\sigma_e$, the real part scales as $1 - \omega^2_{\rmP e}/\omega^2$, just as in the large mean free path limit. Hence, $\varepsilon_{\bk\parallel}(\bk\cdot\bv)$, the dielectric factor by which the acceleration of a charged particle with velocity $\bv$ is modified in the plasma, scales as $1 - \omega^2_{\rmP e}/{\left(\bk\cdot\bv\right)}^2$ for the high $v$ particles, independent of whether the mean free path is small or large compared to the scale of the perturbation, $k^{-1}$. This is because, for these very fast particles, the free-streaming scale $v/\nu_{\rmc e}$ always exceeds $k^{-1}$, even if the mean free path (average free-streaming scale) is small. In other words, the fastest particles are unaffected by collisions.

\begin{figure}[t!]
\centering
\includegraphics[width=0.85\textwidth]{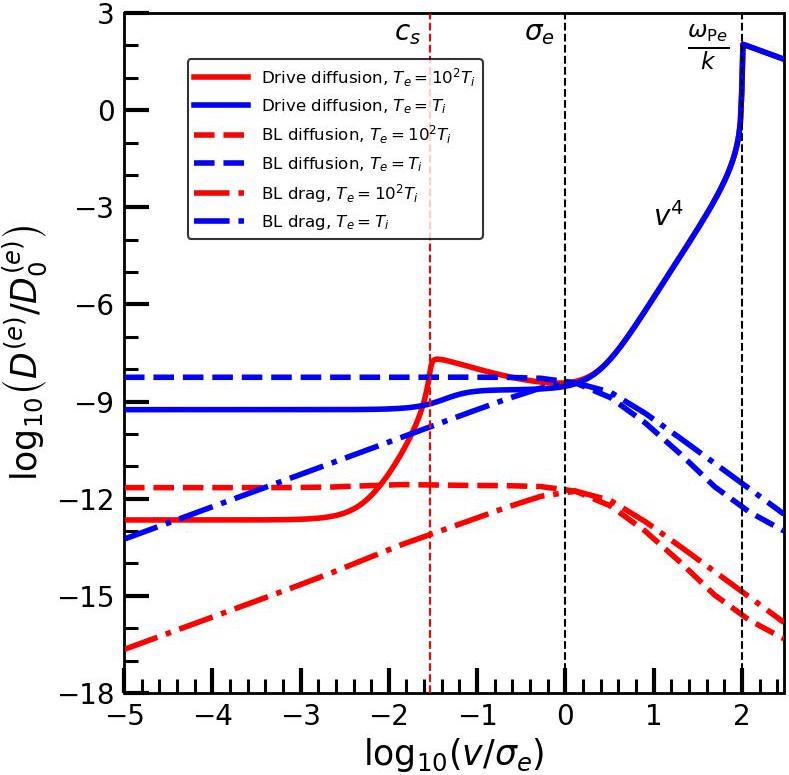}
\caption{Diffusion and drag coefficients vs $v$ for electrons}
\label{fig:D_vs_v}
\end{figure}




\subsection{Quasilinear theory}\label{App:QLT}

In deriving the quasilinear transport coefficients, we have only included the direct acceleration by the drive (as well as internal turbulence) and neglected the indirect heating by the Landau modes or waves. This can be justified as follows. 

Let us first discuss the large mean free path case, $\nu_{\rmc \rms} \ll k\sigma_\rms$. We consider both the electron and ion DFs to be stable to perturbations. On super-Debye scales ($k\lambda_{\rmD} \ll 1$), the electron and ion Langmuir waves damp away at a rate faster than the quasilinear relaxation rate $\sim {\left(k\lambda_\rmD\right)}^4\sigma^2_{\rms}/2 D_{\rms 0}$ ($D_{\rms 0}$ is the diffusion coefficient at $v \gtrsim \omega_{\rmP e}/k$), at which the power-law tail develops at intermediate $v$, by a factor of $\sim {\left(k\lambda_{\rmD}\right)}^{-2}$. The ion-acoustic waves damp away even faster, at a rate $\sim {\left(k\lambda_{\rmD}\right)}^{-3}$ higher than the quasilinear relaxation rate. On sub-Debye ($k\lambda_{\rmD} \gtrsim 1$) scales, Landau damping is efficient and rapidly damps away the waves on a timescale $\sim {\left(k\sigma_\rms\right)}^{-1} \sim {\left(k\lambda_{\rmD}\right)}^{-1}{\omega}^{-1}_{\rmP\rms}$. Therefore, the longitudinal modes do not contribute to quasilinear diffusion at long time. The transverse light waves do not Landau damp away, though, and can potentially contribute. However, for an isotropic plasma subject to isotropic perturbations, the transverse components do not show up in the quasilinear transport equation. Under anisotropic conditions, they do, and the undamped light waves contribute, albeit with a strength suppressed by a factor $\sim {\left(\sum_\rms \omega^2_{\rmP\rms}/c^2k^2\right)}^2$ relative to direct diffusion for sub-skin depth ($k\gtrsim \sqrt{\sum_\rms \omega^2_{\rmP\rms}}/c$) perturbations.

In the small mean free path limit $(\nu_{\rmc \rms} \gg k \sigma_\rms)$, the least damped Landau mode (the Lenard-Bernstein mode \citep[][]{Lenard.Bernstein.58,Banik.Bhattacharjee.24b}), damps at the rate, ${\left(k\sigma_\rms\right)}^2/\nu_{\rmc \rms} \sim {\left(k\lambda_\rmD\right)}^2 \omega_{\rmP\rms}\,\Lambda/\ln\Lambda$. The ratio of this damping rate to the quasilinear diffusion rate is equal to $\sim {\left(k\lambda_\rmD\right)}^{-2} \Lambda/\ln\Lambda$. Hence, the Landau modes damp faster than the power-law tail develops for super-Debye $(k\lambda_\rmD \ll 1)$ perturbations, even more so in a more collisionless environment (larger $\Lambda$). Based on the above considerations, we have neglected the Landau term in the quasilinear transport coefficients \citep[see also equation~(A19) of][]{Banik.Bhattacharjee.24a}.

The quasilinear equation~(\ref{quasilin_resp_FP_eq}) encapsulates particle diffusion due to the drive as well as diffusion and drag due to internal turbulence and collisions. The latter is described by the Balescu-Lenard (BL) diffusion and drag coefficients respectively given by

\begin{align}
&\calD_{ij}^{(\rms)}\left(\bv\right) = \frac{\pi}{m^2_\rms}{\left(4\pi q_\rms q_{\rms'}\right)}^2 \sum_{\rms'} \int \frac{\rmd^3 k}{{\left(2\pi\right)}^3} \frac{k_i k_j}{k^4} \nonumber\\
&\times \frac{1}{{\left|\varepsilon_{\bk\parallel}\left(\bk\cdot\bv\right)\right|}^2} \int \rmd^3 u'\, \delta\left(\bk\cdot\left(\bv-\bv'\right)\right) f_{\rms'0}\left(\bv'\right),\nonumber\\
&\calD_i^{(\rms)}\left(\bv\right) = -\frac{\pi}{m_\rms}{\left(4\pi q_\rms q_{\rms'}\right)}^2  \sum_{\rms'} \frac{1}{m_{\rms'}} \int \frac{\rmd^3 k}{{\left(2\pi\right)}^3} \nonumber\\
&\times \frac{1}{{\left|\varepsilon_{\bk\parallel}\left(\bk\cdot\bv\right)\right|}^2} \int \rmd^3 u'\, \delta\left(\bk\cdot\left(\bv-\bv'\right)\right) \frac{\partial f_{\rms' 0}}{\partial v'_j}.
\label{BL_coeff_app}
\end{align}
Note that only the longitudinal modes contribute to BL diffusion and drag under isotropic conditions in a non-relativistic plasma. In the relativistic regime, the transverse modes also contribute \citep[][]{Silin.61}. The $v$ dependence of the BL coefficients are compared to that of the drive diffusion coefficient in Fig.~\ref{fig:D_vs_v}. Note that, at high enough $v$, a large-scale drive always wins.

\section{Criterion for development of the non-thermal tail}\label{App:criteria}

Under what circumstances does the drive diffusion coefficient exceed the internal coefficient for turbulent relaxation, thereby giving rise to the non-thermal tail? It is easily seen that $D^{(\rms)}\left(v\approx \omega_{\rmP e}/k_\rmc\right)\sim {\left(q_\rms E_{\rm drive}/m_\rms\right)}^2 t_\rmc$ with $E_{\rm drive}$ the strength of the perturbing electric field, $t_\rmc$ its correlation time and $k_\rmc$ its characteristic wavenumber. It is also not hard to see that $\calD^{(\rms)}_2\left(v\approx \omega_{\rmP e}/k_\rmc\right) \sim \sigma^2_\rms\, \omega_{\rmP\rms}\,{\left(\omega_{\rmP e}/k\sigma_\rms\right)}^{-3}\ln\Lambda/\Lambda$, where $\Lambda \sim n_e \lambda^3_{\rmD e}$ is the plasma parameter or the number of electrons within the Debye sphere and $\ln\Lambda$ is the Coulomb logarithm. The strength of the internal electric field is given by $E_{\rm int}$ with $e E_{\rm int}/m_e \sim \omega_{\rmP e}\sigma_e$. For the tail to form, we require that $D^{(\rms)}\left(v\approx \omega_{\rmP e}/k\right) \gtrsim \calD^{(\rms)}_2\left(v\approx \omega_{\rmP e}/k\right)$, i.e.,

\begin{align}
&\frac{E_{\rm drive}}{E_{\rm int}} \gtrsim \delta_\rmc^{(\rms)} \nonumber\\
&={\left(k_\rmc\lambda_{\rmD e}\right)}^{3/2}\sqrt{\frac{\ln\Lambda}{\Lambda} \frac{t_{\rmP e}}{t_\rmc}} \times
\begin{cases}
1,\qquad \qquad \qquad \quad\,\, {\rm electrons},\nonumber\\
Z^{-1/2} {\left(\frac{T_i}{T_e}\right)}^{5/4} {\left(\frac{m_e}{m_i}\right)}^{1/2},\; {\rm ions},
\end{cases}
\end{align}
with $t_{\rmP e} = 2\pi/\omega_{\rm Pe}$ the oscillation period of electron Langmuir waves. Interestingly, the threshold field for the tail is smaller for the ions than for electrons, since the BL diffusion coefficient is smaller for the former. And, the more collisionless the plasma (larger the $\Lambda$), the lower is the threshold field. If $E_{\rm drive}$ crosses this threshold, the core of the distribution below a critical velocity $v_{\rm crit}^{(\rms)}$ Maxwellianizes via turbulent (BL) relaxation and collisions, but particles with $v\gtrsim v_{\rm crit}^{(\rms)}$ are efficiently heated and develop the non-thermal tail. The critical velocity is equal to $\left(\omega_{\rmP e}/k_\rmc\right) {\left(E_{\rm drive}/\delta^{(\rms)}_\rmc E_{\rm int}\right)}^{-2/7}$ if $E_{\rm drive}/\delta_\rmc^{(\rms)} E_{\rm int} > {\left(k_\rmc \lambda_\rmD\right)}^{7/2}$, and ${\rm max}\left(\sigma_e,\nu_{\rmc\rms}/k_\rmc\right)$ otherwise (assuming $T_e\sim T_i$).

\section{Effective EOS from $\kappa$ distributions}\label{App:EOS}
The Vlasov equation dictates that, in a collisionless plasma under the gravitational potential of the sun, both the electron and ion DFs become $f_{\rms 0}(v,r)\sim {\left[1 + {(v^2 + \Phi(r) - \Phi(r_\rmb))}/2\kappa\sigma^2_e\right]}^{-\left(1+\kappa\right)}$, with a particular preference for $\kappa = 1.5$, that is super-exponentially truncated beyond $v_{\rm max} = \omega_{\rmP e}/k_\rmc$. This implies that the total mass density, $\rho(r) = 4\pi \sum_{\rms}\int \rmd v\, v^2 f_{\rms 0}(v,r)$, and the total pressure, $p(r) = 4\pi \sum_{\rms}\int \rmd v\, v^4 f_{\rms 0}(v,r)$, can be written as


\begin{align}
\rho(r) &\sim {\left[1 + {\left(\Phi(r) - \Phi(r_\rmb)\right)}/{2\kappa\sigma^2_e}\right]}^{1/2-\kappa} \beta\left(\sin^2\theta_{\rm max},\frac{3}{2},\kappa-\frac{1}{2}\right),\nonumber\\
p(r) &\sim {\left[1 + {\left(\Phi(r) - \Phi(r_\rmb)\right)}/{2\kappa\sigma^2_e}\right]}^{3/2-\kappa} \beta\left(\sin^2\theta_{\rm max},\frac{5}{2},\kappa-\frac{3}{2}\right),
\label{p_rho_app}
\end{align}
with 

\begin{align}
\theta_{\rm max} = \tan^{-1}\left(\frac{v_{\rm max}}{\sqrt{\left(\Phi(r) - \Phi(r_\rmb) + 2\kappa\sigma^2_e\right)}}\right),
\end{align}
where $v_{\rm max} \approx {\rm min}\left(\omega_{\rmP e}/k_\rmc.1/k_\rmc t_\rmc\right)$, and $\beta\left(z,a,b\right) = \int_0^z \rmd u\, u^{a-1} {\left(1-u\right)}^{b-1}$ is the incomplete beta function. The temperature $T = \left(\mu m_\rmP/k_\rmB\right)\left(p/\rho\right)$ is therefore given by


\begin{align}
T(r) &= T(r_\rmb) + \frac{\mu G M_\odot m_\rmP}{2\kappa k_\rmB} \left(\frac{1}{r_\rmb} - \frac{1}{r}\right)
\end{align}

For $\kappa>3/2$, the effective EOS is a polytropic one: $p\propto \rho^\gamma$ with $\gamma = \left(\kappa-3/2\right)/\left(\kappa-1/2\right)$. For $\kappa = 3/2$, the density and pressure become

\begin{align}
\rho(r) &\sim {\left[1 + {\left(\Phi(r) - \Phi(r_\rmb)\right)}/{3\sigma^2_e}\right]}^{-1}\, \frac{\sin^3\theta_{\rm max}}{3},\nonumber\\
p(r) &\sim \frac{1}{2}\ln{\left|\frac{1 + \sin{\theta_{\rm max}}}{1 - \sin{\theta_{\rm max}}}\right|} - \left(\sin{\theta_{\rm max}} + \frac{\sin^3{\theta_{\rm max}}}{3}\right).
\end{align}
In the limit of $v_{\rm max} \gg \sqrt{\left(\Phi(r) - \Phi(r_\rmb) + 3\sigma^2_e\right)}$, we have that $\sin{\theta_{\rm max}} \approx 1 - \left(\Phi(r) - \Phi(r_\rmb) + 3\sigma^2_e\right)/v^2_{\rm max}$, and the corresponding density scales as $\rho(r)\sim {\left[1 + {\left(\Phi(r) - \Phi(r_\rmb)\right)}/{3\sigma^2_e}\right]}^{-1}$, while the pressure as $p(r)\sim \ln{\left|\left(2 v^2_{\rm max}/3\sigma^2_e\right)\rho(r)/\rho(r_\rmb)\right|}$. This limit amounts to $\rho(r) \gg \left(3\sigma^2_e/v^2_{\rm max}\right)\rho(r_\rmb)$, i.e., near the coronal base where the density is large. In the opposite limit of $v_{\rm max} \ll \sqrt{\left(\Phi(r) - \Phi(r_\rmb) + 3\sigma^2_e\right)}$, $\sin{\theta_{\rm max}} \approx \theta_{\rm max} \approx {v_{\rm max}}/{\sqrt{\left(\Phi(r) - \Phi(r_\rmb) + 3\sigma^2_e\right)}}$, which implies that $p(r)\propto \rho(r) \sim {\left[1 + {\left(\Phi(r) - \Phi(r_\rmb)\right)}/{3\sigma^2_e}\right]}^{-5/2}$. This holds for $\rho(r) \ll \left(3\sigma^2_e/v^2_{\rm max}\right)\rho(r_\rmb)$, i.e., in the low density environment of the outer corona. Thus, for $\kappa = 3/2$, pressure decreases very mildly with density at high density, but linearly with density at low density, implying that $T(r)\sim p(r)/\rho(r)$ steeply increases outwards before flattening out in the nearly isothermal condition of the outer corona.
\\

\section{Entropy considerations}\label{App:entropy}
Can we construct an entropy functional for the effective collision operator in equation~(\ref{QL_eq}), that satisfies the H-theorem (never decreases with time) and yields the above DF as an extremal solution? It can be shown that $S = -\int \rmd^3 x\,\rmd^3v\,G(f)$ with any convex function $G(f)$ follows the H-theorem. Since the DF approaches a $\kappa$ distribution at long times in the presence of a large-scale EM drive, $G(f)\sim \left(f^{q}-1\right)/\left(q-1\right)$ with $q = \kappa/\left(1+\kappa\right)$ is the function for which the entropy $S$, when maximized with the constraints of total energy, momentum and particle number conservation, yields this DF as the extremal solution. The corresponding entropy $S$ is nothing but the Tsallis entropy \citep{Tsallis.88,Livadiotis.McComas.13,Zhdankin.22a,Zhdankin.22b}. And, since our DF scales as $v^{-5}$ at large $v$, $\kappa = 1.5$ and $q = 3/5$ are the preferred values.

Interestingly, \citep[][]{Ewart.etal.22,Ewart.etal.23} recast the entropy into a generalized Boltzmann-Shannon form, $S = \int \rmd \bv \int \rmd \eta\, P(\bv,\eta)\ln{P(\bv,\eta)}$, with $P(\eta)$ the probability that $f$ (treated as a random variable) takes the value $\eta$. Maximizing this with the constraints of the conservation of probability, total energy and phase-volume (waterbag content), $\rho(\eta) = \int \rmd^3v\,P(\bv,\eta) = \frac{1}{V}\int\rmd\bx\int\rmd\bv\,\delta\left(f(\bx,\bv)-\eta\right)$, they obtain a Fermi-Dirac distribution for $P(\bv,\eta)$, along the lines of \citep[][]{LyndenBell.67}. Noting that $\rho(\eta) = \int \rmd \bv\,\delta\left(f_\rmG-\eta\right)$ ($f_\rmG$ is the Gardener distribution that is a function of $E$ and not $\bx$ and has the same $\rho(\eta)$ as $f$) scales as $\eta^{-1}$ for exponentially truncated (or steeper) $f_\rmG$, they find that the Fermi-Dirac $P(\bv,\eta)$ yields $N(E)\sim E^{-2}$ in the non-degenerate limit. It is, however, a priori, not clear why the Gardener distribution would be of such a form. It is easy to see that, for $f_\rmG(v)\sim v^{-\alpha}$, $\rho(\eta)\sim \eta^{-\left(1+1/\alpha\right)}$ (as also pointed out by \citep[][]{Ewart.etal.22}), which would yield a shallower tail than $E^{-2}$. The derivation of the $E^{-2}$ tail from this generalized Lynden-Bell approach hinges on (i) a sufficient deviation of $f_{\rms 0}$ from $f_\rmG$ and (ii) the $\eta^{-1}$ scaling of $\rho(\eta)$. As shown by \citep[][]{Ewart.etal.23} using $\left(1x,1v\right)$ PIC simulations of two-stream instability in electrostatic plasmas, the former depends on the precise nature of Vlasov turbulence (e.g., presence or absence of BGK holes). Interestingly, the $E^{-2}$ tail appears in the spatially averaged energy distribution after the saturation of the two-stream instability, only when BGK holes appear in the phase-space, such as in an electron-ion plasma. The electron-positron case neither features large-scale holes nor the $E^{-2}$ tail. Therefore, it appears that the appearance of the tail is intimately related to the occurrence of a large-scale drive, which, in this case, consists of the electric fields exerted by the BGK holes themselves on the bulk plasma. And the BGK holes are an outcome of the self-consistent plasma response. In other words, self-consistency through the Poisson equation, something that the entropy approach does not explicitly take into account but our kinetic formalism does, is crucial for the emergence of the power-law tail. As we show, it is the Debye-shielding of large-scale EM fields (e.g., those exerted by coherent structures as in \citep[][]{Ewart.etal.23}'s simulation) and the consequent suppression of particle heating at low $v$, that fundamentally gives rise to the tail.


\bibliography{references_banik}

\begin{thebibliography}{32}%
\makeatletter
\providecommand \@ifxundefined [1]{%
 \@ifx{#1\undefined}
}%
\providecommand \@ifnum [1]{%
 \ifnum #1\expandafter \@firstoftwo
 \else \expandafter \@secondoftwo
 \fi
}%
\providecommand \@ifx [1]{%
 \ifx #1\expandafter \@firstoftwo
 \else \expandafter \@secondoftwo
 \fi
}%
\providecommand \natexlab [1]{#1}%
\providecommand \enquote  [1]{``#1''}%
\providecommand \bibnamefont  [1]{#1}%
\providecommand \bibfnamefont [1]{#1}%
\providecommand \citenamefont [1]{#1}%
\providecommand \href@noop [0]{\@secondoftwo}%
\providecommand \href [0]{\begingroup \@sanitize@url \@href}%
\providecommand \@href[1]{\@@startlink{#1}\@@href}%
\providecommand \@@href[1]{\endgroup#1\@@endlink}%
\providecommand \@sanitize@url [0]{\catcode `\\12\catcode `\$12\catcode `\&12\catcode `\#12\catcode `\^12\catcode `\_12\catcode `\%12\relax}%
\providecommand \@@startlink[1]{}%
\providecommand \@@endlink[0]{}%
\providecommand \url  [0]{\begingroup\@sanitize@url \@url }%
\providecommand \@url [1]{\endgroup\@href {#1}{\urlprefix }}%
\providecommand \urlprefix  [0]{URL }%
\providecommand \Eprint [0]{\href }%
\providecommand \doibase [0]{https://doi.org/}%
\providecommand \selectlanguage [0]{\@gobble}%
\providecommand \bibinfo  [0]{\@secondoftwo}%
\providecommand \bibfield  [0]{\@secondoftwo}%
\providecommand \translation [1]{[#1]}%
\providecommand \BibitemOpen [0]{}%
\providecommand \bibitemStop [0]{}%
\providecommand \bibitemNoStop [0]{.\EOS\space}%
\providecommand \EOS [0]{\spacefactor3000\relax}%
\providecommand \BibitemShut  [1]{\csname bibitem#1\endcsname}%
\let\auto@bib@innerbib\@empty
\bibitem [{\citenamefont {Fisk}\ and\ \citenamefont {Gloeckler}(2014)}]{Fisk.Gloeckler.14}%
  \BibitemOpen
  \bibfield  {author} {\bibinfo {author} {\bibfnamefont {L.~A.}\ \bibnamefont {Fisk}}\ and\ \bibinfo {author} {\bibfnamefont {G.}~\bibnamefont {Gloeckler}},\ }\bibfield  {title} {\bibinfo {title} {The case for a common spectrum of particles accelerated in the heliosphere: Observations and theory},\ }\href {https://doi.org/https://doi.org/10.1002/2014JA020426} {\bibfield  {journal} {\bibinfo  {journal} {Journal of Geophysical Research: Space Physics}\ }\textbf {\bibinfo {volume} {119}},\ \bibinfo {pages} {8733} (\bibinfo {year} {2014})}\BibitemShut {NoStop}%
\bibitem [{\citenamefont {{Fisk}}\ and\ \citenamefont {{Gloeckler}}(2012)}]{Fisk.Gloeckler.12}%
  \BibitemOpen
  \bibfield  {author} {\bibinfo {author} {\bibfnamefont {L.~A.}\ \bibnamefont {{Fisk}}}\ and\ \bibinfo {author} {\bibfnamefont {G.}~\bibnamefont {{Gloeckler}}},\ }\bibfield  {title} {\bibinfo {title} {{Particle Acceleration in the Heliosphere: Implications for Astrophysics}},\ }\href {https://doi.org/10.1007/s11214-012-9899-8} {\bibfield  {journal} {\bibinfo  {journal} {\ssr}\ }\textbf {\bibinfo {volume} {173}},\ \bibinfo {pages} {433} (\bibinfo {year} {2012})}\BibitemShut {NoStop}%
\bibitem [{\citenamefont {{Krimigis}}\ \emph {et~al.}(1977)\citenamefont {{Krimigis}}, \citenamefont {{Armstrong}}, \citenamefont {{Axford}}, \citenamefont {{Bostrom}}, \citenamefont {{Fan}}, \citenamefont {{Gloeckler}},\ and\ \citenamefont {{Lanzerotti}}}]{Krimigis.etal.77}%
  \BibitemOpen
  \bibfield  {author} {\bibinfo {author} {\bibfnamefont {S.~M.}\ \bibnamefont {{Krimigis}}}, \bibinfo {author} {\bibfnamefont {T.~P.}\ \bibnamefont {{Armstrong}}}, \bibinfo {author} {\bibfnamefont {W.~I.}\ \bibnamefont {{Axford}}}, \bibinfo {author} {\bibfnamefont {C.~O.}\ \bibnamefont {{Bostrom}}}, \bibinfo {author} {\bibfnamefont {C.~Y.}\ \bibnamefont {{Fan}}}, \bibinfo {author} {\bibfnamefont {G.}~\bibnamefont {{Gloeckler}}},\ and\ \bibinfo {author} {\bibfnamefont {L.~J.}\ \bibnamefont {{Lanzerotti}}},\ }\bibfield  {title} {\bibinfo {title} {{The Low Energy Charged Particle (LECP) Experiment on the Voyager Spacecraft}},\ }\href {https://doi.org/10.1007/BF00211545} {\bibfield  {journal} {\bibinfo  {journal} {\ssr}\ }\textbf {\bibinfo {volume} {21}},\ \bibinfo {pages} {329} (\bibinfo {year} {1977})}\BibitemShut {NoStop}%
\bibitem [{\citenamefont {{Stone}}\ \emph {et~al.}(1977)\citenamefont {{Stone}}, \citenamefont {{Vogt}}, \citenamefont {{McDonald}}, \citenamefont {{Teegarden}}, \citenamefont {{Trainor}}, \citenamefont {{Jokipii}},\ and\ \citenamefont {{Webber}}}]{Stone.etal.77}%
  \BibitemOpen
  \bibfield  {author} {\bibinfo {author} {\bibfnamefont {E.~C.}\ \bibnamefont {{Stone}}}, \bibinfo {author} {\bibfnamefont {R.~E.}\ \bibnamefont {{Vogt}}}, \bibinfo {author} {\bibfnamefont {F.~B.}\ \bibnamefont {{McDonald}}}, \bibinfo {author} {\bibfnamefont {B.~J.}\ \bibnamefont {{Teegarden}}}, \bibinfo {author} {\bibfnamefont {J.~H.}\ \bibnamefont {{Trainor}}}, \bibinfo {author} {\bibfnamefont {J.~R.}\ \bibnamefont {{Jokipii}}},\ and\ \bibinfo {author} {\bibfnamefont {W.~R.}\ \bibnamefont {{Webber}}},\ }\bibfield  {title} {\bibinfo {title} {{Cosmic ray investigation for the Voyager missions; energetic particle studies in the outer heliosphere{\textemdash}And beyond}},\ }\href {https://doi.org/10.1007/BF00211546} {\bibfield  {journal} {\bibinfo  {journal} {\ssr}\ }\textbf {\bibinfo {volume} {21}},\ \bibinfo {pages} {355} (\bibinfo {year} {1977})}\BibitemShut {NoStop}%
\bibitem [{\citenamefont {Jokipii}\ and\ \citenamefont {Lee}(2010)}]{Jokipii.Lee.10}%
  \BibitemOpen
  \bibfield  {author} {\bibinfo {author} {\bibfnamefont {J.~R.}\ \bibnamefont {Jokipii}}\ and\ \bibinfo {author} {\bibfnamefont {M.~A.}\ \bibnamefont {Lee}},\ }\bibfield  {title} {\bibinfo {title} {Compression acceleration in astrophysical plasmas and the production of $f(v) \propto v^{-5}$ spectra in the heliosphere},\ }\href {https://doi.org/10.1088/0004-637X/713/1/475} {\bibfield  {journal} {\bibinfo  {journal} {The Astrophysical Journal}\ }\textbf {\bibinfo {volume} {713}},\ \bibinfo {pages} {475} (\bibinfo {year} {2010})}\BibitemShut {NoStop}%
\bibitem [{\citenamefont {{Ewart}}\ \emph {et~al.}(2025)\citenamefont {{Ewart}}, \citenamefont {{Nastac}}, \citenamefont {{Bilbao}}, \citenamefont {{Silva}}, \citenamefont {{Silva}},\ and\ \citenamefont {{Schekochihin}}}]{Ewart.etal.25}%
  \BibitemOpen
  \bibfield  {author} {\bibinfo {author} {\bibfnamefont {R.~J.}\ \bibnamefont {{Ewart}}}, \bibinfo {author} {\bibfnamefont {M.~L.}\ \bibnamefont {{Nastac}}}, \bibinfo {author} {\bibfnamefont {P.~J.}\ \bibnamefont {{Bilbao}}}, \bibinfo {author} {\bibfnamefont {T.}~\bibnamefont {{Silva}}}, \bibinfo {author} {\bibfnamefont {L.~O.}\ \bibnamefont {{Silva}}},\ and\ \bibinfo {author} {\bibfnamefont {A.~A.}\ \bibnamefont {{Schekochihin}}},\ }\bibfield  {title} {\bibinfo {title} {{Relaxation to universal non-Maxwellian equilibria in a collisionless plasma}},\ }\href {https://doi.org/10.1073/pnas.2417813122} {\bibfield  {journal} {\bibinfo  {journal} {Proceedings of the National Academy of Science}\ }\textbf {\bibinfo {volume} {122}},\ \bibinfo {eid} {e2417813122} (\bibinfo {year} {2025})}\BibitemShut {NoStop}%
\bibitem [{\citenamefont {{Scudder}}(1992)}]{Scudder.92}%
  \BibitemOpen
  \bibfield  {author} {\bibinfo {author} {\bibfnamefont {J.~D.}\ \bibnamefont {{Scudder}}},\ }\bibfield  {title} {\bibinfo {title} {{On the Causes of Temperature Change in Inhomogeneous Low-Density Astrophysical Plasmas}},\ }\href {https://doi.org/10.1086/171858} {\bibfield  {journal} {\bibinfo  {journal} {\apj}\ }\textbf {\bibinfo {volume} {398}},\ \bibinfo {pages} {299} (\bibinfo {year} {1992})}\BibitemShut {NoStop}%
\bibitem [{\citenamefont {{Meyer-Vernet}}(2007)}]{Meyer-Vernet.07}%
  \BibitemOpen
  \bibfield  {author} {\bibinfo {author} {\bibfnamefont {N.}~\bibnamefont {{Meyer-Vernet}}},\ }\href {https://doi.org/10.1017/CBO9780511535765} {\emph {\bibinfo {title} {{Basics of the Solar Wind}}}}\ (\bibinfo {year} {2007})\BibitemShut {NoStop}%
\bibitem [{\citenamefont {{Anderson}}(1994)}]{Anderson.94}%
  \BibitemOpen
  \bibfield  {author} {\bibinfo {author} {\bibfnamefont {S.~W.}\ \bibnamefont {{Anderson}}},\ }\bibfield  {title} {\bibinfo {title} {{Coulomb Collisions and Coronal Heating by Velocity Filtration}},\ }\href {https://doi.org/10.1086/175046} {\bibfield  {journal} {\bibinfo  {journal} {\apj}\ }\textbf {\bibinfo {volume} {437}},\ \bibinfo {pages} {860} (\bibinfo {year} {1994})}\BibitemShut {NoStop}%
\bibitem [{\citenamefont {{Landi}}\ and\ \citenamefont {{Pantellini}}(2001)}]{Landi.Pantellini.01}%
  \BibitemOpen
  \bibfield  {author} {\bibinfo {author} {\bibfnamefont {S.}~\bibnamefont {{Landi}}}\ and\ \bibinfo {author} {\bibfnamefont {F.~G.~E.}\ \bibnamefont {{Pantellini}}},\ }\bibfield  {title} {\bibinfo {title} {{On the temperature profile and heat flux in the solar corona: Kinetic simulations}},\ }\href {https://doi.org/10.1051/0004-6361:20010552} {\bibfield  {journal} {\bibinfo  {journal} {\aap}\ }\textbf {\bibinfo {volume} {372}},\ \bibinfo {pages} {686} (\bibinfo {year} {2001})}\BibitemShut {NoStop}%
\bibitem [{\citenamefont {Diamond}\ \emph {et~al.}(2010)\citenamefont {Diamond}, \citenamefont {Itoh},\ and\ \citenamefont {Itoh}}]{Diamond_Itoh_Itoh_2010}%
  \BibitemOpen
  \bibfield  {author} {\bibinfo {author} {\bibfnamefont {P.~H.}\ \bibnamefont {Diamond}}, \bibinfo {author} {\bibfnamefont {S.-I.}\ \bibnamefont {Itoh}},\ and\ \bibinfo {author} {\bibfnamefont {K.}~\bibnamefont {Itoh}},\ }\href@noop {} {\emph {\bibinfo {title} {Modern Plasma Physics}}}\ (\bibinfo  {publisher} {Cambridge University Press},\ \bibinfo {year} {2010})\BibitemShut {NoStop}%
\bibitem [{\citenamefont {{Bernstein}}\ \emph {et~al.}(1957)\citenamefont {{Bernstein}}, \citenamefont {{Greene}},\ and\ \citenamefont {{Kruskal}}}]{Bernstein.etal.57}%
  \BibitemOpen
  \bibfield  {author} {\bibinfo {author} {\bibfnamefont {I.~B.}\ \bibnamefont {{Bernstein}}}, \bibinfo {author} {\bibfnamefont {J.~M.}\ \bibnamefont {{Greene}}},\ and\ \bibinfo {author} {\bibfnamefont {M.~D.}\ \bibnamefont {{Kruskal}}},\ }\bibfield  {title} {\bibinfo {title} {{Exact Nonlinear Plasma Oscillations}},\ }\href {https://doi.org/10.1103/PhysRev.108.546} {\bibfield  {journal} {\bibinfo  {journal} {Physical Review}\ }\textbf {\bibinfo {volume} {108}},\ \bibinfo {pages} {546} (\bibinfo {year} {1957})}\BibitemShut {NoStop}%
\bibitem [{\citenamefont {{Loureiro}}\ \emph {et~al.}(2007)\citenamefont {{Loureiro}}, \citenamefont {{Schekochihin}},\ and\ \citenamefont {{Cowley}}}]{Loureiro.etal.07}%
  \BibitemOpen
  \bibfield  {author} {\bibinfo {author} {\bibfnamefont {N.~F.}\ \bibnamefont {{Loureiro}}}, \bibinfo {author} {\bibfnamefont {A.~A.}\ \bibnamefont {{Schekochihin}}},\ and\ \bibinfo {author} {\bibfnamefont {S.~C.}\ \bibnamefont {{Cowley}}},\ }\bibfield  {title} {\bibinfo {title} {{Instability of current sheets and formation of plasmoid chains}},\ }\href {https://doi.org/10.1063/1.2783986} {\bibfield  {journal} {\bibinfo  {journal} {Physics of Plasmas}\ }\textbf {\bibinfo {volume} {14}},\ \bibinfo {pages} {100703} (\bibinfo {year} {2007})},\ \Eprint {https://arxiv.org/abs/astro-ph/0703631} {arXiv:astro-ph/0703631 [astro-ph]} \BibitemShut {NoStop}%
\bibitem [{\citenamefont {{Bhattacharjee}}\ \emph {et~al.}(2009)\citenamefont {{Bhattacharjee}}, \citenamefont {{Huang}}, \citenamefont {{Yang}},\ and\ \citenamefont {{Rogers}}}]{Bhattacharjee.etal.09}%
  \BibitemOpen
  \bibfield  {author} {\bibinfo {author} {\bibfnamefont {A.}~\bibnamefont {{Bhattacharjee}}}, \bibinfo {author} {\bibfnamefont {Y.-M.}\ \bibnamefont {{Huang}}}, \bibinfo {author} {\bibfnamefont {H.}~\bibnamefont {{Yang}}},\ and\ \bibinfo {author} {\bibfnamefont {B.}~\bibnamefont {{Rogers}}},\ }\bibfield  {title} {\bibinfo {title} {{Fast reconnection in high-Lundquist-number plasmas due to the plasmoid Instability}},\ }\href {https://doi.org/10.1063/1.3264103} {\bibfield  {journal} {\bibinfo  {journal} {Physics of Plasmas}\ }\textbf {\bibinfo {volume} {16}},\ \bibinfo {eid} {112102} (\bibinfo {year} {2009})},\ \Eprint {https://arxiv.org/abs/0906.5599} {arXiv:0906.5599 [physics.plasm-ph]} \BibitemShut {NoStop}%
\bibitem [{\citenamefont {{Comisso}}\ \emph {et~al.}(2016)\citenamefont {{Comisso}}, \citenamefont {{Lingam}}, \citenamefont {{Huang}},\ and\ \citenamefont {{Bhattacharjee}}}]{Comisso.etal.16}%
  \BibitemOpen
  \bibfield  {author} {\bibinfo {author} {\bibfnamefont {L.}~\bibnamefont {{Comisso}}}, \bibinfo {author} {\bibfnamefont {M.}~\bibnamefont {{Lingam}}}, \bibinfo {author} {\bibfnamefont {Y.~M.}\ \bibnamefont {{Huang}}},\ and\ \bibinfo {author} {\bibfnamefont {A.}~\bibnamefont {{Bhattacharjee}}},\ }\bibfield  {title} {\bibinfo {title} {{General theory of the plasmoid instability}},\ }\href {https://doi.org/10.1063/1.4964481} {\bibfield  {journal} {\bibinfo  {journal} {Physics of Plasmas}\ }\textbf {\bibinfo {volume} {23}},\ \bibinfo {eid} {100702} (\bibinfo {year} {2016})},\ \Eprint {https://arxiv.org/abs/1608.04692} {arXiv:1608.04692 [physics.plasm-ph]} \BibitemShut {NoStop}%
\bibitem [{\citenamefont {{Banik}}\ \emph {et~al.}(2024)\citenamefont {{Banik}}, \citenamefont {{Bhattacharjee}},\ and\ \citenamefont {{Sengupta}}}]{Banik.Bhattacharjee.24a}%
  \BibitemOpen
  \bibfield  {author} {\bibinfo {author} {\bibfnamefont {U.}~\bibnamefont {{Banik}}}, \bibinfo {author} {\bibfnamefont {A.}~\bibnamefont {{Bhattacharjee}}},\ and\ \bibinfo {author} {\bibfnamefont {W.}~\bibnamefont {{Sengupta}}},\ }\bibfield  {title} {\bibinfo {title} {{Universal Nonthermal Power-law Distribution Functions from the Self-consistent Evolution of Collisionless Electrostatic Plasmas}},\ }\href {https://doi.org/10.3847/1538-4357/ad91a1} {\bibfield  {journal} {\bibinfo  {journal} {\apj}\ }\textbf {\bibinfo {volume} {977}},\ \bibinfo {eid} {91} (\bibinfo {year} {2024})},\ \Eprint {https://arxiv.org/abs/2408.07127} {arXiv:2408.07127 [astro-ph.SR]} \BibitemShut {NoStop}%
\bibitem [{\citenamefont {Fermi}(1949)}]{Fermi.49}%
  \BibitemOpen
  \bibfield  {author} {\bibinfo {author} {\bibfnamefont {E.}~\bibnamefont {Fermi}},\ }\bibfield  {title} {\bibinfo {title} {On the origin of the cosmic radiation},\ }\href {https://doi.org/10.1103/PhysRev.75.1169} {\bibfield  {journal} {\bibinfo  {journal} {Phys. Rev.}\ }\textbf {\bibinfo {volume} {75}},\ \bibinfo {pages} {1169} (\bibinfo {year} {1949})}\BibitemShut {NoStop}%
\bibitem [{\citenamefont {{Nastac}}\ \emph {et~al.}(2023)\citenamefont {{Nastac}}, \citenamefont {{Ewart}}, \citenamefont {{Sengupta}}, \citenamefont {{Schekochihin}}, \citenamefont {{Barnes}},\ and\ \citenamefont {{Dorland}}}]{Nastac.etal.23}%
  \BibitemOpen
  \bibfield  {author} {\bibinfo {author} {\bibfnamefont {M.~L.}\ \bibnamefont {{Nastac}}}, \bibinfo {author} {\bibfnamefont {R.~J.}\ \bibnamefont {{Ewart}}}, \bibinfo {author} {\bibfnamefont {W.}~\bibnamefont {{Sengupta}}}, \bibinfo {author} {\bibfnamefont {A.~A.}\ \bibnamefont {{Schekochihin}}}, \bibinfo {author} {\bibfnamefont {M.}~\bibnamefont {{Barnes}}},\ and\ \bibinfo {author} {\bibfnamefont {W.~D.}\ \bibnamefont {{Dorland}}},\ }\bibfield  {title} {\bibinfo {title} {{Phase-space entropy cascade and irreversibility of stochastic heating in nearly collisionless plasma turbulence}},\ }\href {https://doi.org/10.48550/arXiv.2310.18211} {\bibfield  {journal} {\bibinfo  {journal} {arXiv e-prints}\ ,\ \bibinfo {eid} {arXiv:2310.18211}} (\bibinfo {year} {2023})},\ \Eprint {https://arxiv.org/abs/2310.18211} {arXiv:2310.18211 [physics.plasm-ph]} \BibitemShut {NoStop}%
\bibitem [{\citenamefont {{Nastac}}\ \emph {et~al.}(2025)\citenamefont {{Nastac}}, \citenamefont {{Ewart}}, \citenamefont {{Juno}}, \citenamefont {{Barnes}},\ and\ \citenamefont {{Schekochihin}}}]{Nastac.etal.25}%
  \BibitemOpen
  \bibfield  {author} {\bibinfo {author} {\bibfnamefont {M.~L.}\ \bibnamefont {{Nastac}}}, \bibinfo {author} {\bibfnamefont {R.~J.}\ \bibnamefont {{Ewart}}}, \bibinfo {author} {\bibfnamefont {J.}~\bibnamefont {{Juno}}}, \bibinfo {author} {\bibfnamefont {M.}~\bibnamefont {{Barnes}}},\ and\ \bibinfo {author} {\bibfnamefont {A.~A.}\ \bibnamefont {{Schekochihin}}},\ }\bibfield  {title} {\bibinfo {title} {{Universal fluctuation spectrum of Vlasov-Poisson turbulence}},\ }\href {https://doi.org/10.48550/arXiv.2503.17278} {\bibfield  {journal} {\bibinfo  {journal} {arXiv e-prints}\ ,\ \bibinfo {eid} {arXiv:2503.17278}} (\bibinfo {year} {2025})},\ \Eprint {https://arxiv.org/abs/2503.17278} {arXiv:2503.17278 [physics.plasm-ph]} \BibitemShut {NoStop}%
\bibitem [{\citenamefont {{Ginat}}\ \emph {et~al.}(2025)\citenamefont {{Ginat}}, \citenamefont {{Nastac}}, \citenamefont {{Ewart}}, \citenamefont {{Konrad}}, \citenamefont {{Bartelmann}},\ and\ \citenamefont {{Schekochihin}}}]{Ginat.etal.25}%
  \BibitemOpen
  \bibfield  {author} {\bibinfo {author} {\bibfnamefont {Y.~B.}\ \bibnamefont {{Ginat}}}, \bibinfo {author} {\bibfnamefont {M.~L.}\ \bibnamefont {{Nastac}}}, \bibinfo {author} {\bibfnamefont {R.~J.}\ \bibnamefont {{Ewart}}}, \bibinfo {author} {\bibfnamefont {S.}~\bibnamefont {{Konrad}}}, \bibinfo {author} {\bibfnamefont {M.}~\bibnamefont {{Bartelmann}}},\ and\ \bibinfo {author} {\bibfnamefont {A.~A.}\ \bibnamefont {{Schekochihin}}},\ }\bibfield  {title} {\bibinfo {title} {{Gravitational Turbulence: the Small-Scale Limit of the Cold-Dark-Matter Power Spectrum}},\ }\href {https://doi.org/10.48550/arXiv.2501.01524} {\bibfield  {journal} {\bibinfo  {journal} {arXiv e-prints}\ ,\ \bibinfo {eid} {arXiv:2501.01524}} (\bibinfo {year} {2025})},\ \Eprint {https://arxiv.org/abs/2501.01524} {arXiv:2501.01524 [astro-ph.CO]} \BibitemShut {NoStop}%
\bibitem [{\citenamefont {{Livadiotis}}\ and\ \citenamefont {{McComas}}(2013)}]{Livadiotis.McComas.13}%
  \BibitemOpen
  \bibfield  {author} {\bibinfo {author} {\bibfnamefont {G.}~\bibnamefont {{Livadiotis}}}\ and\ \bibinfo {author} {\bibfnamefont {D.~J.}\ \bibnamefont {{McComas}}},\ }\bibfield  {title} {\bibinfo {title} {{Understanding Kappa Distributions: A Toolbox for Space Science and Astrophysics}},\ }\href {https://doi.org/10.1007/s11214-013-9982-9} {\bibfield  {journal} {\bibinfo  {journal} {\ssr}\ }\textbf {\bibinfo {volume} {175}},\ \bibinfo {pages} {183} (\bibinfo {year} {2013})}\BibitemShut {NoStop}%
\bibitem [{\citenamefont {{Zhdankin}}(2022{\natexlab{a}})}]{Zhdankin.22a}%
  \BibitemOpen
  \bibfield  {author} {\bibinfo {author} {\bibfnamefont {V.}~\bibnamefont {{Zhdankin}}},\ }\bibfield  {title} {\bibinfo {title} {{Generalized Entropy Production in Collisionless Plasma Flows and Turbulence}},\ }\href {https://doi.org/10.1103/PhysRevX.12.031011} {\bibfield  {journal} {\bibinfo  {journal} {Physical Review X}\ }\textbf {\bibinfo {volume} {12}},\ \bibinfo {eid} {031011} (\bibinfo {year} {2022}{\natexlab{a}})},\ \Eprint {https://arxiv.org/abs/2110.07025} {arXiv:2110.07025 [astro-ph.HE]} \BibitemShut {NoStop}%
\bibitem [{\citenamefont {{Zhdankin}}(2022{\natexlab{b}})}]{Zhdankin.22b}%
  \BibitemOpen
  \bibfield  {author} {\bibinfo {author} {\bibfnamefont {V.}~\bibnamefont {{Zhdankin}}},\ }\bibfield  {title} {\bibinfo {title} {{Non-thermal particle acceleration from maximum entropy in collisionless plasmas}},\ }\href {https://doi.org/10.1017/S0022377822000551} {\bibfield  {journal} {\bibinfo  {journal} {Journal of Plasma Physics}\ }\textbf {\bibinfo {volume} {88}},\ \bibinfo {eid} {175880303} (\bibinfo {year} {2022}{\natexlab{b}})},\ \Eprint {https://arxiv.org/abs/2203.13054} {arXiv:2203.13054 [astro-ph.HE]} \BibitemShut {NoStop}%
\bibitem [{\citenamefont {{Ewart}}\ \emph {et~al.}(2022)\citenamefont {{Ewart}}, \citenamefont {{Brown}}, \citenamefont {{Adkins}},\ and\ \citenamefont {{Schekochihin}}}]{Ewart.etal.22}%
  \BibitemOpen
  \bibfield  {author} {\bibinfo {author} {\bibfnamefont {R.~J.}\ \bibnamefont {{Ewart}}}, \bibinfo {author} {\bibfnamefont {A.}~\bibnamefont {{Brown}}}, \bibinfo {author} {\bibfnamefont {T.}~\bibnamefont {{Adkins}}},\ and\ \bibinfo {author} {\bibfnamefont {A.~A.}\ \bibnamefont {{Schekochihin}}},\ }\bibfield  {title} {\bibinfo {title} {{Collisionless relaxation of a Lynden-Bell plasma}},\ }\href {https://doi.org/10.1017/S0022377822000782} {\bibfield  {journal} {\bibinfo  {journal} {Journal of Plasma Physics}\ }\textbf {\bibinfo {volume} {88}},\ \bibinfo {eid} {925880501} (\bibinfo {year} {2022})},\ \Eprint {https://arxiv.org/abs/2201.03376} {arXiv:2201.03376 [physics.plasm-ph]} \BibitemShut {NoStop}%
\bibitem [{\citenamefont {{Ewart}}\ \emph {et~al.}(2023)\citenamefont {{Ewart}}, \citenamefont {{Nastac}},\ and\ \citenamefont {{Schekochihin}}}]{Ewart.etal.23}%
  \BibitemOpen
  \bibfield  {author} {\bibinfo {author} {\bibfnamefont {R.~J.}\ \bibnamefont {{Ewart}}}, \bibinfo {author} {\bibfnamefont {M.~L.}\ \bibnamefont {{Nastac}}},\ and\ \bibinfo {author} {\bibfnamefont {A.~A.}\ \bibnamefont {{Schekochihin}}},\ }\bibfield  {title} {\bibinfo {title} {{Non-thermal particle acceleration and power-law tails via relaxation to universal Lynden-Bell equilibria}},\ }\href {https://doi.org/10.1017/S0022377823000983} {\bibfield  {journal} {\bibinfo  {journal} {Journal of Plasma Physics}\ }\textbf {\bibinfo {volume} {89}},\ \bibinfo {eid} {905890516} (\bibinfo {year} {2023})},\ \Eprint {https://arxiv.org/abs/2304.03715} {arXiv:2304.03715 [physics.plasm-ph]} \BibitemShut {NoStop}%
\bibitem [{\citenamefont {{Pierrard}}\ and\ \citenamefont {{Lazar}}(2010)}]{Pierrard.Lazar.10}%
  \BibitemOpen
  \bibfield  {author} {\bibinfo {author} {\bibfnamefont {V.}~\bibnamefont {{Pierrard}}}\ and\ \bibinfo {author} {\bibfnamefont {M.}~\bibnamefont {{Lazar}}},\ }\bibfield  {title} {\bibinfo {title} {{Kappa Distributions: Theory and Applications in Space Plasmas}},\ }\href {https://doi.org/10.1007/s11207-010-9640-2} {\bibfield  {journal} {\bibinfo  {journal} {\solphys}\ }\textbf {\bibinfo {volume} {267}},\ \bibinfo {pages} {153} (\bibinfo {year} {2010})},\ \Eprint {https://arxiv.org/abs/1003.3532} {arXiv:1003.3532 [physics.space-ph]} \BibitemShut {NoStop}%
\bibitem [{\citenamefont {{Parker}}(1965)}]{Parker.65}%
  \BibitemOpen
  \bibfield  {author} {\bibinfo {author} {\bibfnamefont {E.~N.}\ \bibnamefont {{Parker}}},\ }\bibfield  {title} {\bibinfo {title} {{The passage of energetic charged particles through interplanetary space}},\ }\href {https://doi.org/10.1016/0032-0633(65)90131-5} {\bibfield  {journal} {\bibinfo  {journal} {\planss}\ }\textbf {\bibinfo {volume} {13}},\ \bibinfo {pages} {9} (\bibinfo {year} {1965})}\BibitemShut {NoStop}%
\bibitem [{\citenamefont {{Lenard}}\ and\ \citenamefont {{Bernstein}}(1958)}]{Lenard.Bernstein.58}%
  \BibitemOpen
  \bibfield  {author} {\bibinfo {author} {\bibfnamefont {A.}~\bibnamefont {{Lenard}}}\ and\ \bibinfo {author} {\bibfnamefont {I.~B.}\ \bibnamefont {{Bernstein}}},\ }\bibfield  {title} {\bibinfo {title} {{Plasma Oscillations with Diffusion in Velocity Space}},\ }\href {https://doi.org/10.1103/PhysRev.112.1456} {\bibfield  {journal} {\bibinfo  {journal} {Physical Review}\ }\textbf {\bibinfo {volume} {112}},\ \bibinfo {pages} {1456} (\bibinfo {year} {1958})}\BibitemShut {NoStop}%
\bibitem [{\citenamefont {{Banik}}\ and\ \citenamefont {{Bhattacharjee}}(2024)}]{Banik.Bhattacharjee.24b}%
  \BibitemOpen
  \bibfield  {author} {\bibinfo {author} {\bibfnamefont {U.}~\bibnamefont {{Banik}}}\ and\ \bibinfo {author} {\bibfnamefont {A.}~\bibnamefont {{Bhattacharjee}}},\ }\bibfield  {title} {\bibinfo {title} {{Relaxation of weakly collisional plasma: Continuous spectra, discrete eigenmodes, and the decay of echoes}},\ }\href {https://doi.org/10.1103/PhysRevE.110.045204} {\bibfield  {journal} {\bibinfo  {journal} {\pre}\ }\textbf {\bibinfo {volume} {110}},\ \bibinfo {eid} {045204} (\bibinfo {year} {2024})},\ \Eprint {https://arxiv.org/abs/2402.07992} {arXiv:2402.07992 [physics.plasm-ph]} \BibitemShut {NoStop}%
\bibitem [{\citenamefont {Silin}(1961)}]{Silin.61}%
  \BibitemOpen
  \bibfield  {author} {\bibinfo {author} {\bibfnamefont {V.~P.}\ \bibnamefont {Silin}},\ }\bibfield  {title} {\bibinfo {title} {On the collision integral for charged particles},\ }\href {https://www.osti.gov/biblio/4836334} {\bibfield  {journal} {\bibinfo  {journal} {Zhur. Eksptl'. i Teoret. Fiz.}\ }\textbf {\bibinfo {volume} {Vol: 40}} (\bibinfo {year} {1961})}\BibitemShut {NoStop}%
\bibitem [{\citenamefont {{Tsallis}}(1988)}]{Tsallis.88}%
  \BibitemOpen
  \bibfield  {author} {\bibinfo {author} {\bibfnamefont {C.}~\bibnamefont {{Tsallis}}},\ }\bibfield  {title} {\bibinfo {title} {{Possible generalization of Boltzmann-Gibbs statistics}},\ }\href {https://doi.org/10.1007/BF01016429} {\bibfield  {journal} {\bibinfo  {journal} {Journal of Statistical Physics}\ }\textbf {\bibinfo {volume} {52}},\ \bibinfo {pages} {479} (\bibinfo {year} {1988})}\BibitemShut {NoStop}%
\bibitem [{\citenamefont {{Lynden-Bell}}(1967)}]{LyndenBell.67}%
  \BibitemOpen
  \bibfield  {author} {\bibinfo {author} {\bibfnamefont {D.}~\bibnamefont {{Lynden-Bell}}},\ }\bibfield  {title} {\bibinfo {title} {{Statistical mechanics of violent relaxation in stellar systems}},\ }\href {https://doi.org/10.1093/mnras/136.1.101} {\bibfield  {journal} {\bibinfo  {journal} {\mnras}\ }\textbf {\bibinfo {volume} {136}},\ \bibinfo {pages} {101} (\bibinfo {year} {1967})}\BibitemShut {NoStop}%
\end{thebibliography}%

\end{document}